\documentclass[reprint,aps,prd,superscriptaddress,showkeys,showpacs]{revtex4-1}
\usepackage{epsfig,amsmath,natbib}
\usepackage{aas_macros}
\usepackage{amssymb}
\usepackage{amsmath}
\usepackage{dsfont}
\usepackage{hyperref}
\usepackage{color}
\usepackage{pbox}
\usepackage{enumerate}

\hypersetup{
	colorlinks=true,
	linkcolor=blue,
	urlcolor=blue,
	citecolor=blue
}
\newcommand\lsim{\mathrel{\rlap{\lower4pt\hbox{\hskip0pt$\sim$}}
        \raise1pt\hbox{$<$}}}
\newcommand\gsim{\mathrel{\rlap{\lower4pt\hbox{\hskip0pt$\sim$}}
        \raise1pt\hbox{$>$}}}

\begin{document}

\title{Probing gaseous galactic halos through the rotational kSZ effect}

\author{Jos\'e Manuel Zorrilla Matilla}
\email{jzorrilla@astro.columbia.edu}
\affiliation{Department of Astronomy, Columbia University, New York, NY 10027, USA}
\author{Zolt\'an Haiman}
\affiliation{Department of Astronomy, Columbia University, New York, NY 10027, USA}

\date{\today}

\begin{abstract}
  The rotational kinematic Sunyaev-Zeldovich (rkSZ) signal, imprinted
  on the cosmic microwave background (CMB) by the gaseous halos
  (spinning ``atmospheres'') of foreground galaxies, would be a novel
  probe of galaxy formation.  Although the signal is too weak to
  detect in individual galaxies, we analyze the feasibility of its
  statistical detection via stacking CMB data on many galaxies for
  which the spin orientation can be estimated spectroscopically.  We
  use an ``optimistic'' model, in which fully ionized atmospheres
  contain the cosmic baryon fraction and spin at the halo's circular
  velocity $v_{\rm circ}$, and a more realistic model, based on
  hydrodynamical simulations, with multi-phase atmospheres spinning at
  a fraction of $v_{\rm circ}$.  We incorporate realistic noise
  estimates into our analysis.
  Using low-redshift galaxy properties from the MaNGA spectroscopic
  survey (with median halo mass of $6.6\times10^{11}\,M_{\odot}$), and
  CMB data quality from \textit{Planck}, we find that a $3\sigma$
  detection would require a few$\times 10^4$ galaxies, even in the
  optimistic model. This is too high for current surveys, but
  upcoming higher-angular resolution CMB experiments will significantly
  reduce the requirements: stacking CMB data on galaxy spins in a $\sim10$~deg$^2$ can rule out the optimistic models, and
  $\approx$350\,deg$^2$ will suffice for a $3\sigma$ detection with
  ACT.
  As a proof-of-concept, we stacked \textit{Planck} data on the
  position of $\approx2,000$ MaNGA galaxies, aligned with the
  galaxies' projected spin, and scaled to their halos' angular size.
  We rule out average temperature dipoles larger than $\approx1.9\,\mu$K around field spiral galaxies.
\end{abstract}
\keywords{Cosmic microwave background, large scale structure of the universe, formation \& evolution of stars \& galaxies.}
\pacs{}
\maketitle

\section{Introduction}\label{rksz1.introduction}

A deeper understanding of galaxy formation and evolution requires
comparing the expected properties of the circumgalactic medium (CGM)
with observations. These properties, predicted by simulations, include
the CGM's density, composition, ionization state, and kinematics, as
well as the evolution of these quantities through cosmic time. For
reviews of the CGM and its connection to galaxy evolution, see,
e.g. \cite{Tumlinson2017,Peeples2019}.

Cosmic microwave background (CMB) photons interact with the free
electrons in the CGM plasma, and can therefore probe the CGM's
properties. The kinematic Sunyaev-Zeldovich effect (kSZ) is the
gain/loss of momentum of these photons as they scatter coherently off
electrons with a bulk motion relative to the CMB
\citep{Sunyaev72}. The kSZ effect can be used to either learn about
the free electron distribution given some kinematic information, or to
infer the CGM's peculiar velocity given its free electron density. We
refer the reader to \citep{Birkinshaw99} for a detailed review of this
effect.

The use of mean pairwise statistics enabled the early detection of the
kSZ effect induced by the proper motions of galaxy clusters
\citep{Hand12}, and the same method has recently been applied
successfully to galaxies \citep{DeBernardis17,Li18}. The kSZ signal
due to clusters' proper motions has also been detected in stacked data
\citep{Lavaux13, Schaan16} and through high-resolution imaging of
individual systems \citep{Adam17}. It has also been detected in
cross-correlation analyses of projected fields \citep{Hill2016,
  Ferraro2016}.

Rotating gaseous halos should imprint an additional, dipole-like
temperature pattern in the CMB at their location. This signal, which 
we will refer to as rotational kSZ effect (rkSZ, as in \citep{Baxter19}), 
appears
on small angular scales ($\lsim 10$~arcmin, corresponding to the
halo virial radii $R_{\rm vir}$), and has been
studied in the context of galaxy clusters both
analytically~\citep{Cooray02, Chluba02} and with simulations
\citep{Baldi18}. Recently, a tentative detection has been claimed
\citep{Baxter19}, stacking \textit{Planck}
data \footnote{\textit{Planck}
  collaboration:https://www.cosmos.esa.int/web/planck} on the location
of rotating clusters identified in the Sloan Digital Sky Survey
(SDSS, \footnote{Sloan Digital Sky
  Survey:\url{https://www.sdss.org/}}).

In the near future, high-resolution CMB experiments will allow an
extension of these studies to probe the rotation of the gaseous halos
of individual galaxies.  While the signal-to-noise for individual
galaxies will remain too low, spin orientations can be estimated for
large numbers of nearby galaxies in forthcoming spectroscopic surveys.
Motivated by this prospect, in this paper, we
assess the feasibility of detecting the rkSZ effect via
stacking CMB data on many galaxies. At present, asymmetries in the CMB
temperature aligned with the rotation axis of nearby galaxies have
been measured \citep{DePaolis2014, DePaolis2016, Gurzadyan2015,
  Gurzadyan2018}, but the origin of these asymmetries is not yet fully
understood (see \S~\ref{rksz1.discussion}).

Our manuscript is organized as follows. We start with a description of
our models for the rkSZ signal from gaseous galactic
halos~(\S~\ref{rksz1.model}).  We next describe how to stack CMB data
and extract the rkSZ signal
statistically~(\S~\ref{rksz1.stats}), and forecast the number of
galaxies needed for a $3\sigma$ detection for a variety of
experimental settings~(\S~\ref{rksz1.forecast}). We then proceed to
apply these techniques to existing public CMB and galaxy survey data
(\S~\ref{rksz1.mangaplanck}).  In particular, we derive an upper limit
on the mean CMB temperature asymmetry in \textit{Planck} data,
associated with galaxy spins in the spectroscopic MaNGA survey
(Mapping Nearby Galaxies at APO, \footnote{Mapping Nearby Galaxies at
  APO:\url{https://www.sdss.org/surveys/manga/}}). Finally, we discuss
different caveats and extensions of our analysis and results (\S~
\ref{rksz1.discussion}) and summarize our main
conclusions~(\S~\ref{rksz1.conclusion}).

All calculations assume a flat $\Lambda$CDM cosmology with
$\Omega_m=0.316$, $\Omega_b=0.048$, $h=0.675$ and T$_{\rm
  cmb}=2.725\,$K.

\section{Modeling the rotational kSZ (rkSZ) signal from galaxies} \label{rksz1.model}

\subsection{The  rkSZ imprint on the CMB} \label{rksz1.model.ksz}

Free electrons moving relative to the Hubble flow induce temperature anisotropies on the CMB through scattering. This kinetic effect is frequency-independent and cannot be isolated from the primordial CMB in the same way as the thermal SZ effect. The kSZ-induced temperature fluctuations depend on the line-of-sight ({\it los}) integral of the density as well as the peculiar velocity of the electrons. Since the CGM is optically thin to photons from the CMB, we can express the relative change in temperature from free electrons in a galactic halo using the single-scattering limit,

\begin{equation}\label{eq:rksz1.model.ksz.ksz}
\begin{split}
    \frac{\Delta T}{T}\left(\Vec{n}\right)=\frac{\sigma_T}{c}\int_{los} dl \, n_e \Vec{v}\cdot \Vec{n}\\ = \frac{\sigma_T}{c}\int_{los} dl \, n_e\left(r\right) v\left(R\right) \cos{\phi} \sin{i}
\end{split}
\end{equation}

where $\Vec{n}$ is the unit vector that defines the point on the sky where the CMB temperature is measured, $\sigma_T$ is the Thomson cross section, $c$ is the speed of light, $n_e$ is the electron density, and $\Vec{v}$ is the velocity of the electrons in the CMB rest frame. The last equality applies to a spherically symmetric distribution of free electron ($r$ is the distance to the halo's center), moving along circular orbits of radius $R$ with velocity $v\left(R\right)$. The azimuth angle is $\phi$ and the galaxy's inclination angle $i$ (0\,deg for a face-on galaxy, 90\,deg for edge-on). Fig.~\ref{fig:rksz1.model.signal} shows an example of the dipole-like temperature anisotropy induced by a rotating halo on the CMB.

For simplicity, we do not include the kSZ effect induced by the galaxy's mean peculiar velocity in our models, but we discuss its effect on measurements in \S~\ref{rksz1.discussion.modellimitations} (along with the effect of uncertainties on model parameters, such as the inclination angle and stellar mass of each galaxy).

\subsection{Galactic atmospheres: electron density} \label{rksz1.model.ne}

The first ingredient needed to estimate the rkSZ signal is the electron density, which, for simplicity, we assume to be spherically symmetric, $n_e(r)$. 

A simple reference model, which has been used in the study of rotating galaxy clusters, is one with fully ionized hot gas in hydrostatic equilibrium within the gravitational potential of the galaxy's host dark matter (DM) halo \citep{Cooray02, Chluba02, Baldi18, Baxter19}. While this model cannot describe galaxy-sized halos, for which a significant fraction of the gas is in a cold and neutral phase, it is still useful as an upper limit to the electron number density. We will refer to such a galactic atmosphere as ``hot".

\begin{figure}
    \centering
    \includegraphics[width=\linewidth]{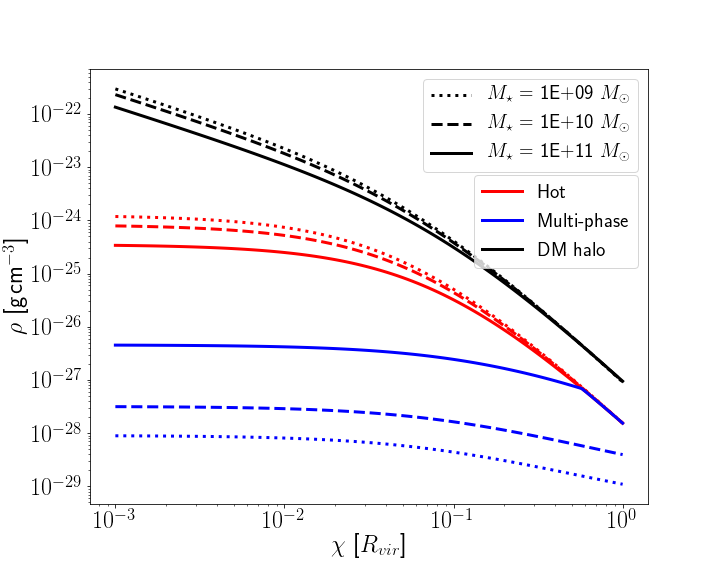}
    \caption{Density profiles for the ionized gas for fully ionized galactic atmospheres (``hot", in red) and multi-phase atmospheres (``multi-phase", in blue) as a function of distance to the center in virial radii units for galaxies of three different stellar masses. The gas metallicity is $Z_g=0.3~{\rm Z_\odot}$ and the DM density profile of the host halo is shown in black for reference. The halos' virial masses are $\{1.2\times10^{11}, 3.4\times10^{11}, 3.3\times10^{12}\}$\,$M_{\odot}$, their virial radii $\{69, 98, 215\}$\,kpc, and their concentrations $\{17.3, 15.0, 11.0\}$, respectively.}
    \label{fig:rksz1.model.ne.gasprofile}
\end{figure}

A more realistic electron density distribution is given by the multi-phase atmospheric model developed in \citep{Maller04}. We reproduce here its main equations for convenience, and refer the reader to \citep{Maller04} for more detailed explanations. The difference in the distribution of ionized gas in this model and the hot upper limit is shown in Fig.~\ref{fig:rksz1.model.ne.gasprofile} for three galaxies of different mass. 

The starting point for the multi-phase model is the galaxy's stellar mass, which we assume to be independent of redshift, that is $M_{\star}(z)\approx M_{\star}(0)$. This approximation is justified because we only work with galaxies in the local universe. For a given $M_{\star}(0)$, we find the virial mass of the galaxy's host halo, $M_v(0)$, using the fit in \citep{Kravtsov18} (see Eqs.\,A3-A4 in their appendix). We then scale $M_v(0)$ to the halo mass $M_v(z)$ at earlier redshift using the relationship, based on $N$-body simulations, in \citep{Wechsler02}:

\begin{equation}\label{eq:rksz1.model.ne.haloaccretion}
    M_v(z)=M_v(0)\exp{\left(-\frac{8.2 z}{C_v^0}\right)}.
\end{equation}

This assumes an NFW profile \citep{NFW97} for the host DM halos. The mass-dependent NFW halo concentration parameter at zero redshift, $C_v^0$, is derived from the fit to simulations in \citep{Bullock01}:

\begin{equation}\label{eq:rksz1.model.ne.haloconcentration}
    C_v^0 = 9 \left(\frac{M_v(0)}{1.5 \times 10^{13}/ h M_{\odot}}\right)^{-0.13}.
\end{equation}

We define the halo's virial radius and mass following the equations in \citep{Bryan98}.  For simplicity, we further assume that the total baryonic mass inside a halo corresponds to its cosmic mass fraction, $f_b=\Omega_b / \Omega_m$.  We relax this assumption, and discuss how lower baryon fractions affect our results, in \S~\ref{rksz1.discussion.modellimitations}.

The hot atmosphere, used as an upper bound, is fully determined by the DM halo mass and its baryon fraction (see Eqs.\,9-11 in \citep{Maller04}). Defining $\xi\equiv r/r_s$ as the dimensionless radial coordinate normalized by the halo's scale radius $r_s\equiv R_{\rm vir}/C_v^0$, the free electron density profile for the hot atmosphere, $n_e^h(\xi)$, is given by:

\begin{eqnarray}\label{eq:rksz1.model.ne.hotgas}
    n_e^h(\xi)=\frac{\rho_0}{\mu_e m_p \left(\xi + \frac{3}{4} \right)\left(\xi+1\right)^2},\\
    \rho_0 = \frac{f_b M_v}{4\pi r_s^3 g(C_v)}, \\
    g(x)\equiv 9 \ln{\left(1+\frac{4}{3}x\right)}-8\ln{\left(1+x\right)}-\frac{4x}{1+x}.
\end{eqnarray}

We use a mean atomic weight per electron $\mu_e=1.18$ (appropriate for ionized gas with mean cosmological abundance ratios) and $m_p$ is the proton mass.

In galaxy-sized halos, a significant fraction of the baryons cool and condense into a neutral phase, a part of which form stars. The cooling time of the gas depends on its density, temperature, and cooling rate, $\Lambda$. For a halo whose time since its formation is $t_f$, the electron density above which hot gas has had time to cool is:
\begin{equation}\label{eq:rksz1.model.ne.cooling}
    n_e^c = \frac{3\mu_e k_b T}{2 \mu_i t_f \Lambda(T, Z_g)},
\end{equation}
where $k_b$ is the Boltzmann constant, $T$ is the temperature corresponding to the halo's maximum circular velocity, and $Z_g$ the metallicity of the gas. The halo's formation time is the lookback time to the redshift at which it has accreted half its mass. We adopt the cooling function parametrized in Appendix A of \citep{Maller04} for a metallicity of $Z_g=0.3~{\rm Z_\odot}$.

The density in the outer regions of massive halos is below $n_e^c$, gas hasn't had time to cool, and the free electron density is given by Eq.~\ref{eq:rksz1.model.ne.hotgas}. In the inner regions, the density exceeds $n_e^c$ and most gas cools into a neutral phase. The transition between the two regimes takes place at the cooling radius. In the inner regions, there will still be some residual hot gas. We will refer to the ionized component of galactic atmospheres as ``coronae". Assuming the hot corona reaches hydrostatic equilibrium adiabatically and its density at the cooling radius matches the cooling density, its free electron density is given by:
\begin{equation}\label{eq:rksz1.model.ne.hotcorona}
    n_e^{hc}(\xi) = n_e^c\left[]1 + \frac{3.7}{\xi}\ln{\left(1+\xi\right)}-\frac{3.7}{\xi_c}\ln{\left(1+\xi_c\right)}\right]^{3/2},
\end{equation}
where $\xi_c$ is the dimensionless cooling radius. We will refer to this, more realistic, galactic atmosphere model as ``multi-phase".

\subsection{Galactic atmosphere: kinematics} \label{rksz1.model.kin}

The second ingredient needed is the velocity field of the free electrons. As for the free electron density, we consider two models: an upper bound and a more realistic rotational velocity. In both cases, we assume the velocity field has cylindrical symmetry. 

The upper bound model presumes baryons rotate at the host halo's circular velocity, which can be expressed as a function of the cylindrical radial coordinate normalized by the halo's scale radius, $\varrho$:
\begin{equation}\label{eq:rksz1.model.kin.circularvel}
    v_c(\varrho) = \sqrt{\frac{4 \pi G r_s^2\left[\ln{\left(1+\varrho\right)}-
    \varrho(1+\varrho)^{-1}\right]}{\varrho}}.
\end{equation}

The angular momentum of galaxies and their halos is typically expressed in terms of the ratio between the system's angular velocity and the one corresponding to full rotational support, or spin parameter, $\lambda$. While for DM halos $\lambda$ is generally small ($\lambda \approx 0.05$, see \cite{Bullock2001momentum}), for gas it can reach order unity when it collapses towards the halo's center as it cools and is observed in disk galaxies. We refer to a model with $\lambda=1$, whose circular velocity is given by Eq.~\ref{eq:rksz1.model.kin.circularvel} as ``fast", or a ``fast rotator".

\begin{figure}
    \centering
    \includegraphics[width=\linewidth]{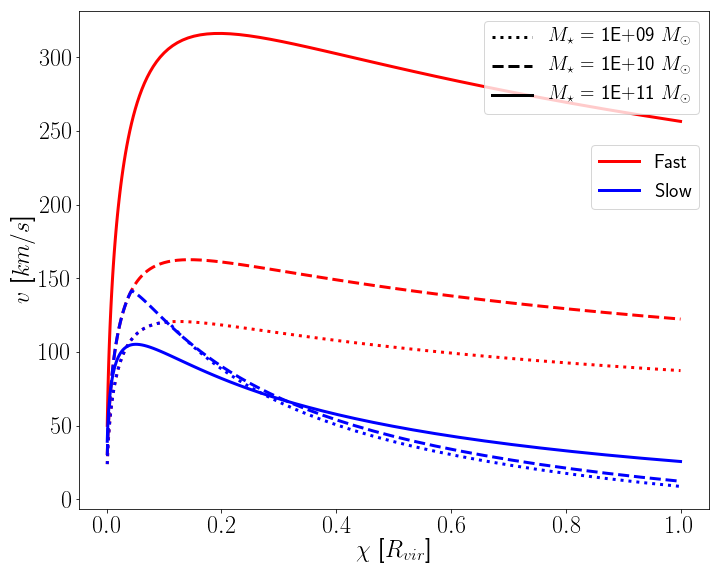}
    \caption{Radial profiles of the tangential gas velocity for a fast rotator (in red, rotating at the halo's circular velocity) and a slow rotator (in blue, rotating at a velocity consistent with simulations~\citep{Oppenheimer18}) for galaxies of three different stellar masses.}
    \label{fig:rksz1.model.kin.velprofile}
\end{figure}

We also consider a more realistic model with $\lambda <1$, and define its velocity field as a fraction of the circular velocity: $v(\varrho)=f(\varrho, M_v) v_c(\varrho)$. This fraction depends on the halo's mass and the distance to its center. We use measurements of the tangential velocity of hot gas in hydrodynamical simulations (see Fig.\,3 in \citep{Oppenheimer18}) to determine $f$. For low-mass halos ($\lsim 10^{13}\,M_{\odot}$), this velocity drops from $\approx 75\%$ of the virial velocity (defined as the circular velocity at the virial radius) in the inner regions to $\approx 10\%$ at the virial radius. For high-mass halos ($\gsim 10^{13}\,M_{\odot}$), the ratio of velocities remains roughly constant at $\approx 10\%$. Instead of the virial velocity, we use the circular velocity as a normalization, to avoid $\lambda>1$ in the halos' innermost regions. We use as a fitting formula:

\begin{equation}\label{eq:rksz1.model.kin.fit1}
    f = \min\left\{m \log{\frac{R}{R_{\rm vir}}}+0.1, 1\right\},
\end{equation}
\begin{equation}\label{eq:rksz1.model.kin.fit2}
    m = -6.0\times10^{-6} \left(\log{M_v}\right)^2 + 3.2\times10^{-1} \log{M_v} -4.4.
\end{equation}
The resulting velocity profile, which we will refer to as ``slow",
is shown in Fig~\ref{fig:rksz1.model.kin.velprofile}, for three different halo masses, together with the alternative, "fast" profiles. As the figure shows, the "slow" profiles are much less sensitive to the halo's mass. The velocities derived from the slow model are in agreement with those predicted for Milky Way and M31 analogs (see \citep{Nuza14}).

\begin{figure}
    \centering
    \includegraphics[width=\linewidth]{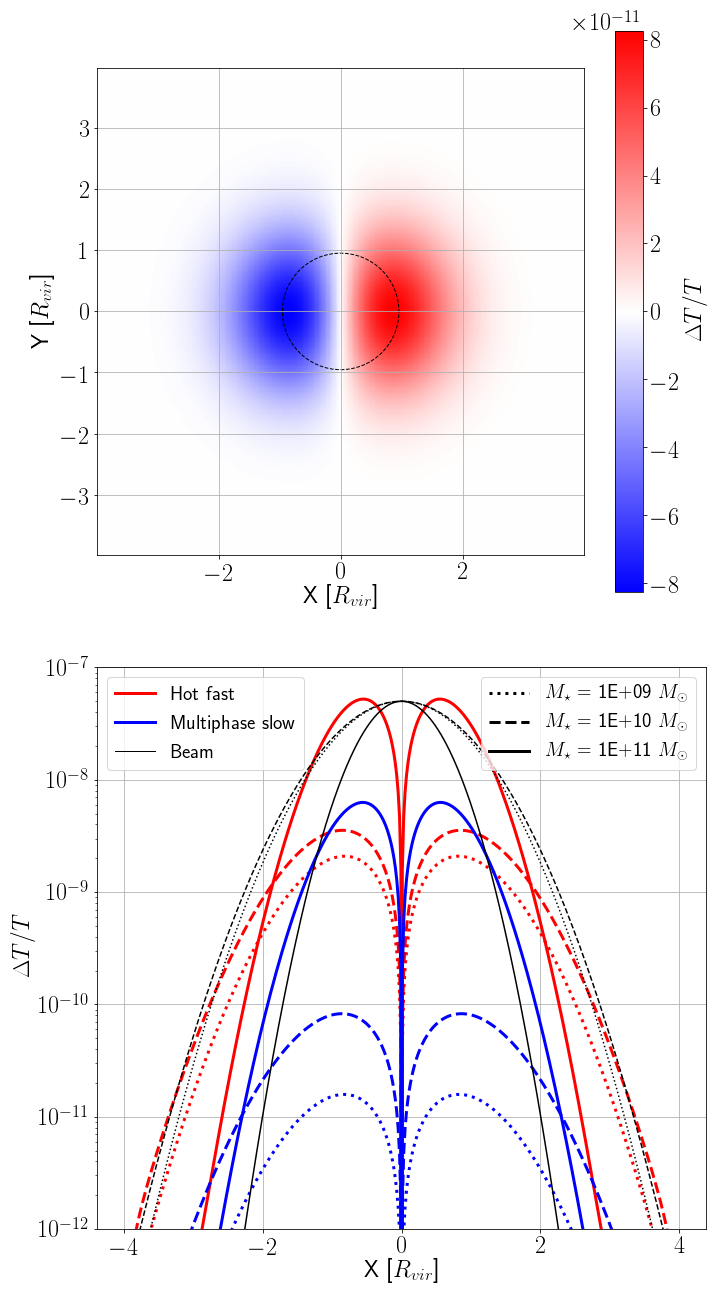}
    \caption{\textbf{Upper panel:} 2D map of the expected fractional temperature change induced in the CMB by the rotation of a $M_*=10^{10}\,M_{\odot}$ galaxy hosted by a $3.4\times 10^{11}\,M_{\odot}$ DM halo with a virial radius of 98.1\,kpc and a concentration parameter of $C_{v}=15.0$ at a redshift of $z=0.03$, assuming the multi-phase slow model. The galaxy's inclination is 1\,rad and the signal has been convolved with a 5\,arcmin FWHM beam (represented by the small dotted circles at the center), while the halo virial radius has an apparent size of 5.2\,arcmin.\\
    \textbf{Lower panel:} cut along the X-axis of the dipole-like signal in the upper panel, for galaxies of three different stellar masses. The predictions in the hot+fast model (fully ionized atmosphere rotating at the halo's circular velocity) are shown in red, and the multi-phase, slow model are shown in blue. The signal was calculated for a metallicity of $Z_g=0.3~{\rm Z_\odot}$, and shown in absolute value. The profile of the convolving beam is displayed in black.
    }
    \label{fig:rksz1.model.signal}
\end{figure}

\section{Characterizing the observed rkSZ signal} \label{rksz1.stats}

Detecting the dipole-like kSZ signal induced by the rotation of galactic halos is challenging, compared to that from galaxy clusters. The signal is diminished by the smaller projected electron number density (due to smaller halos), the lower ionization fraction (due to some of the gas cooling and recombining), and smaller angular size on the sky (the beam width for a given CMB experiment will smooth the signal). Stacking the signal from many galaxies is then a necessity.

Different spatial filters can be used to extract the signal from the noise in the stacked data. In this study, we consider two filters: an aperture filter that measures the temperature asymmetry between its right and left halves, and a matched filter designed based on the profile of the expected signal.  We next discuss these filters, as well as the expected noise levels and resulting signal-to-noise ratios in both cases.

\subsection{Aperture filter}\label{rksz1.stats.aperture}

An aperture filter that measures the temperature difference between its two halves can be used to measure a rkSZ signal, as long as it is centered on the galaxies' and its halves aligned with the galaxies' projected spin vector. A simple statistic is the mean temperature on the right minus the left half of the filter (or dipole):
\begin{equation}\label{eq:rksz1.stats.aperture.dipole}
    s\equiv\overline{\Delta T}^R - \overline{\Delta T}^L.
\end{equation}
In the absence of a rkSZ effect, we expect this statistic to average to zero. It is a robust statistic, in the sense that it is sensitive to any CMB temperature asymmetry relative to the galaxies' projected spin vectors, regardless of the specific shape of the asymmetry. It is also insensitive to any symmetric (on average) signal induced by the halos, such as the thermal SZ effect (tSZ) or the kSZ effect due to the galaxies' peculiar velocities.

Even in the absence of any kSZ effect, a dipole may arise due to random anisotropies in the CMB within the aperture filter. While the mean dipole due to the CMB's random fluctuations is zero ($\langle s \rangle_{\rm cmb}=0$), its variance is not, and should be accounted for as noise. The variance is sourced by both CMB temperature anisotropies and by instrumental noise. Combining both contributions in a single angular power spectrum $C_{\ell}=C_{\ell}^{\rm cmb} + C_{\ell}^{\rm noise}$, the variance of $s$ can be computed as (see \cite{Ferraro2015} and Appendix~\ref{rksz1.appendix.aperturevariance}):
\begin{equation}\label{eq:rksz1.stats.aperture.variance}
\begin{split}
    \sigma^2_s = \left< \left(\overline{\Delta T}^R\right)^2\right> + \left< \left(\overline{\Delta T}^L\right)^2\right> - 2 \left<\overline{\Delta T}^R \overline{\Delta T}^L \right> \\ 
    = 2\left[ \left< \left(\overline{\Delta T}^R\right)^2\right> - \left<\overline{\Delta T}^R \overline{\Delta T}^L \right> \right]
\end{split}
\end{equation}

The aperture filter is defined by its window function $W(x,y)$ in a coordinate system in which the galaxies' spin is aligned with the $y$-axis. The covariance between the mean temperatures measured over the two halves of the window function by an instrument whose beam profile in Fourier space, or beam function, is $b_{\ell}$, can be estimated by
\begin{equation}\label{eq:rksz1.stats.aperture.correlation}
    \left< \overline{\Delta T}^R \overline{\Delta T}^L \right> = 
    \int \frac{\mathrm{d}^2 \boldsymbol{\ell}}{\left(2\pi\right)^2} b_{\ell}^2 C_{\ell} \widetilde{W}_L^{\ast}(\boldsymbol{\ell}) \widetilde{W}_R(\boldsymbol{\ell}),
\end{equation}
where $\widetilde{W}^{\ast}$ is the Fourier transform of $W(x,y)$.  The variance in each of the two halves follows from the analogous expressions, but with
$|\widetilde{W}_L|^2$ or $|\widetilde{W}_R|^2$ in the integrand. The window function for the aperture filter is semi-analytic in Fourier space (see Eq.\ref{eq:rksz1.appendix.aperturevariance.analyticwindow}).

\subsection{Matched filter}\label{rksz1.stats.matched}
The aperture filter described in \S~\ref{rksz1.stats.aperture} is robust, but not optimal, since not all the information encoded in the shape of the signal is used. The optimal approach would be to use a matched filter (see, for example, \citep{Haehnelt1996}). For each galaxy, the optimal filter is essentially the expected rkSZ signal pattern, with the different angular scales weighted by the expected noise (CMB anisotropies and instrumental noise). In Fourier space,
\begin{equation}\label{eq:rksz1.stats.matched.filter}
    \widetilde{\mathrm{MF}}(\boldsymbol{\ell})= \frac{1}{\int \mathrm{d}^2\boldsymbol{\ell} \frac{\left|\widetilde{\Delta T}_{kSZ}(\boldsymbol{\ell})\right|^2}{C_{\ell}}}\frac{\widetilde{\Delta T}_{kSZ}^{\ast}(\boldsymbol{\ell})}{C_{\ell}}.
\end{equation}
For each galaxy, this filter can be applied to the corresponding CMB data, and the result stacked for all galaxies in the survey. The expected signal will make itself apparent as a high peak at the center of the stack. In the absence of any signal, the filtered data will yield pure noise. The height of the central peak relative to the standard deviation in the absence of signal can be used as an estimate for the signal-to-noise ratio (SNR) of the detection via this approach.

Note that additionally, a matched filter is optimal only if the model used for its design corresponds to the true signal in the data. Also, contrary to the aperture filter, the matched filter is not insensitive to potential isotropic signals induced by galactic halos, such as the tSZ effect or kSZ effect induced by peculiar velocities (see \S~\ref{rksz1.discussion.modellimitations}).

\section{Measurement signal-to-noise and required number of galaxies} \label{rksz1.forecast}
In order to coherently stack CMB data for each galaxy, without nulling their rkSZ signal, we need to align the CMB data with each galaxy's projected spin angle. These angles can be measured, for instance, from spatially-resolved spectroscopic data. Integral field spectroscopy enables the efficient acquisition of such data for thousands of galaxies. Examples of recent and ongoing surveys include MaNGA~\footnote{Mapping Nearby Galaxies at APO:\url{https://www.sdss.org/surveys/manga/}} and SAMI~\footnote{Sydney-Australian-Astronomical-Observatory Multi-object Integral-Field Spectrograph: \url{https://sami-survey.org/}}. 

To assess the viability of measuring the stacked rkSZ signal, we estimated the number of galaxies needed for a $3\sigma$ detection.

\begin{figure}
    \centering
    \includegraphics[width=\linewidth]{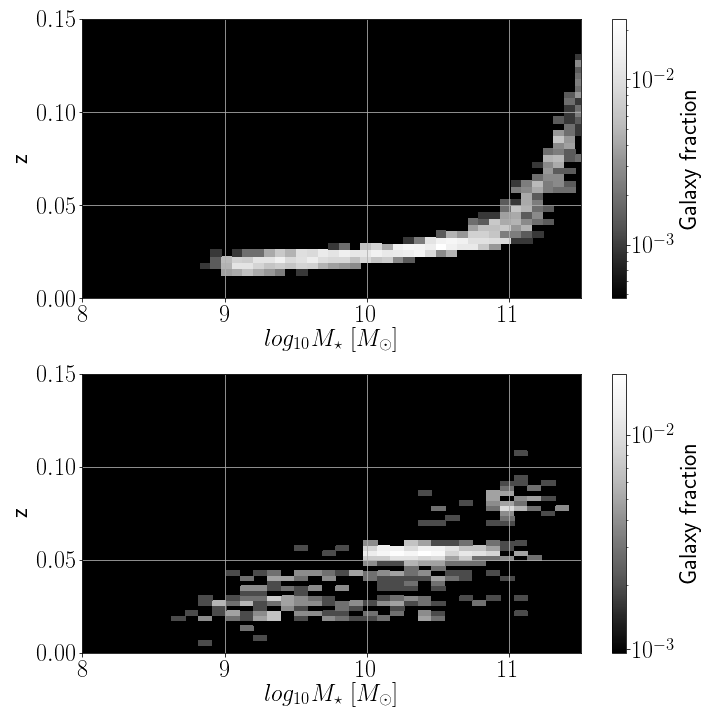}
    \caption{The distribution of galaxies in redshift and stellar mass, in the two prototype surveys considered to assess the feasibility of detecting the kSZ signal induced by the rotation of galactic halos. The top panel corresponds to the primary sample in MaNGA, and the lower panel to SAMI.}
    \label{fig:rksz1.forecast.surveys}
\end{figure}

We considered galaxy surveys with the same redshift and stellar-mass distributions as MaNGA (specifically its "primary" sample, see \S~\ref{rksz1.mangaplanck.manga}) and SAMI. We divided each survey's range of redshift and stellar mass in a $40\times40$ grid, resulting in 235 non-empty bins for MaNGA and 358 for SAMI, shown in Figure~\ref{fig:rksz1.forecast.surveys}. 
The expected signal, $s_i$, and noise, $\sigma_i$, contributed by each bin is estimated from the mean redshift $\langle z_i\rangle$ and stellar mass, $\langle M_{\star i}\rangle$ of galaxies in each bin, for simplicity.

Assuming galaxies are randomly oriented, the probability density function of their inclination angle $i$ is $p(i)=\sin i$, and the mean inclination is $\langle i \rangle=1$~rad, which is the value we adopted for all bins. The SNR for the full survey is a weighted average of the signal and noise in each bin:
\begin{equation}\label{eq:rksz1.forecast.sn}
    \mathrm{SNR}=\sqrt{N_{\rm gal}}\frac{\sum_{i=1}^{N_{\rm bin}} w_i f_i s_i}{\sqrt{\sum_{i=1}^{N_{\rm bin}} w_i^2 f_i \sigma_i^2}}\equiv\sqrt{N_{\rm gal}}\mathrm{SNR_1}.
\end{equation}
Here $N_{\rm gal}$ is the total number of galaxies in the survey, distributed among $N_{\rm bin}$ bins, each with a fraction of the total $f_i$, and SNR$_1$ is the equivalent single-galaxy SNR, which depends on the average properties of the survey's galaxies. After some algebra, it can be shown that the weights that maximize the survey's SNR are $w_i=s_i/\sigma_i^2$.

Future high-resolution CMB experiments will be able to resolve the halos of nearby galaxies. For relatively large scales ($\ell < 10^3$), primordial CMB fluctuations are the dominant source of noise. At smaller scales, we also consider instrumental noise for the five different experimental configurations listed in Table~\ref{tab:rksz1.forecast.cmbexperiments}. Each experiment is characterized by its beam's FWHM and its instrumental noise, which is defined by a white, $\ell$-independent power spectrum \citep{Knox1995}.

The first configuration in Table~\ref{tab:rksz1.forecast.cmbexperiments} corresponds to \textit{Planck} (we use as a reference its 217\,GHz channel, whose frequency is close to the one at which the tSZ is null). The second is the 148\,GHz channel from ACT (the Atacama Cosmology Telescope~\footnote{Atacama Cosmology Telescope:\url{https://act.princeton.edu/}}), the third is the ``Goal" target for the 145\,GHz channel of the Simons observatory (Simons observatory~\footnote{Simons Observatory:\url{https://simonsobservatory.org/}}), the fourth is a possible high-frequency channel of a CMB stage 4 experiment (CMB-S4~\footnote{CMB-S4:\url{https://cmb-s4.org/}}) and the fifth a potential high-resolution future CMB experiment, (CMB-HD \citep{Sehgal2019}).

\begin{table}
\centering
\begin{tabular}{l c c c}
                && FWHM     & $\Delta T_{\rm noise}$   \\
Experiment      && [arcmin] & [$\mu K$\,arcmin]    \\
\hline
\textit{Planck} && 5.00      & 45.6                \\
ACT             && 1.40      & 15.0                \\
Simons          && 1.40      & 6.0                 \\
CMB-S4          && 1.40      & 1.0                 \\
CMB-HD          && 0.25      & 0.5                 \\
\hline
\hline
\end{tabular}
\caption{Instrumental configurations considered for different existing and planned CMB experiments, defined in each case by the beam's FWHM and the rms noise level.}
\label{tab:rksz1.forecast.cmbexperiments}
\end{table}

\begin{figure}
    \centering
    \includegraphics[width=\linewidth]{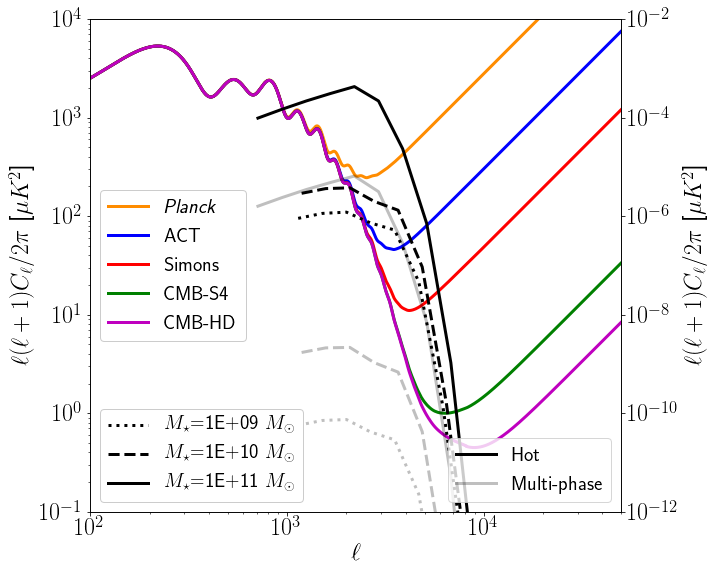}
    \caption{Comparison between the power spectrum of CMB temperature anisotropies, including instrumental noise which dominates at $\ell \gsim$ few$\times10^3$, and that for the expected rkSZ signal for three galaxies of different stellar mass (the same galaxies used in Figs.~\ref{fig:rksz1.model.ne.gasprofile}-\ref{fig:rksz1.model.signal}). 
    The scale on the left y-axis refers to the CMB+noise and the scale on the right to the kSZ power spectra. Note that the rkSZ power is several orders of magnitude lower than that from the combination of CMB + instrumental noise.}
    \label{fig:rksz1.forecast.powerspectrum}
\end{figure}

For illustration, in Figure~\ref{fig:rksz1.forecast.powerspectrum} we show  a comparison of the power spectrum of the intrinsic temperature anisotropies of the CMB (computed with {\tt CAMB} \citep{Lewis1999}) and of the instrumental noise for all five configurations.

Also shown in the figure, for reference, is the power spectrum of the rkSZ signal, $|\Delta T_{kSZ}(\boldsymbol{\ell})|^2$, for galaxies with three different stellar masses (defined simply as the 2D Fourier transform of the signal in Eq.~\ref{eq:rksz1.model.ksz.ksz} and shown in Fig.~\ref{fig:rksz1.model.signal}). The labels on the $y$ axis on the left correspond to the CMB and the instrumental noise, and on the right to the kSZ signal.  The large difference in magnitudes is indicative of the large number of galaxies that will need to be stacked to required to separate the signal from the noise.

We used the two filters described in \S~\ref{rksz1.stats} to compute the signal and the noise contributed by each galaxy bin. The measured signal results from applying the filters to the expected theoretical kSZ signal from the two models detailed in \S~\ref{rksz1.model}: a hot, fast-rotating and a multi-phase, slow-rotating galactic atmosphere. The signal for the aperture filter is the magnitude of the measured temperature dipole, given in Eq.~\ref{eq:rksz1.stats.aperture.dipole}, whereas for the matched filter, it is the height of the central peak of its convolution with the expected theoretical signal, as discussed in \S~\ref{rksz1.stats.matched}.

The noise level for the aperture filter is computed directly from Eq.~\eqref{eq:rksz1.stats.aperture.variance}, using the numerical $C_\ell$ that includes both the primary CMB and the instrumental noise (shown in Fig.~\ref{fig:rksz1.forecast.powerspectrum}).  For the matched filter, in each redshift and stellar-mass bin~$i$, we created 100 independent realizations of synthetic noise-only CMB maps. Each synthetic noise map is generated from a Gaussian random field, defined again by the combined power spectrum of the CMB and the experiment's instrumental noise.  Each map is then convolved with the matched filter for the mean redshift $\langle z_i\rangle$ and stellar mass, $\langle M_{\star i}\rangle$ in that bin and yields a peak height in real space; the noise is computed as the standard deviation of these 100 peak-height values. This exercise yields the signal-to-noise ratio per galaxy SNR$_1$ in Eq.~(\ref{eq:rksz1.forecast.sn}), and the
number of galaxies $N_{\rm gal}$ required for a $3\sigma$ detection follows by setting the total SNR=3
in this equation. 

\begin{figure}
    \centering
    \includegraphics[width=\linewidth]{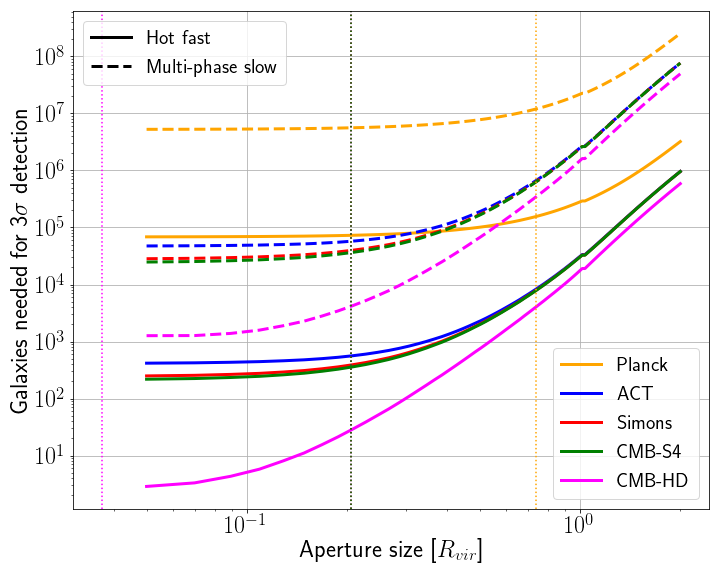}
    \caption{Number of galaxies needed for a $3\sigma$ detection of the rkSZ signal using an aperture filter, as a function of filter size. Each CMB experiment configuration is displayed in a different color. The galaxy $(z, M_{\star})$ distribution is assumed to be that of the primary sample of MaNGA. Solid lines correspond to predictions based on the hot+fast model, while dashed lines are based on the multi-phase, slow model. For reference, vertical lines represent the CMB experiment beam size, in units of the average $R_{\rm vir}$ for the galaxies in the survey.}
    \label{fig:detection.manga}
\end{figure}

Fig.~\ref{fig:detection.manga} shows the number of galaxies required for a $3\sigma$ detection using the aperture filter, as a function of the filter's size, for the set of CMB experiments under consideration. It assumes a MaNGA-like survey (the results for a SAMI-like survey are qualitatively the same with slightly higher number requirements). For a given experiment, the number of galaxies needed decreases as the size of the aperture filter gets smaller, up to the point where the aperture can resolve the galaxies' halos. The resolution of CMB experiments is the main factor that determines their ability to measure the rkSZ signal, over the level of instrumental noise.

A summary of the number of galaxies needed, for both surveys, is shown for each CMB experiment configuration in Table~\ref{tab:rksz1.forecast.results}. The reference aperture size is 0.1\,$R_{\rm vir}$. As Fig.~\ref{fig:detection.manga} shows, the number of galaxies required decreases dramatically as the resolution of the CMB experiments improves. The higher redshift of SAMI galaxies translates into a somewhat larger number of galaxies needed for a detection. The table also shows that matched filtering can reduce the number of galaxies needed, compared with the aperture filter, by a factor of $\approx 2$ for \textit{Planck}, and by more than an order of magnitude for a CMB Stage 4 experiment. However, this statistic is more sensitive to modeling errors, filter misplacements, and isotropic signals on the scale of galactic halos (tSZ, kSZ from peculiar velocities).

While the numbers are large for all the cases, future CMB experiments will be able to rule out most models and may be able to make a detection (see further discussion of the detection feasibility in \S~\ref{rksz1.discussion.feasability} below).

\begin{table}
\centering
\begin{tabular}{l c c c c c}
           && MaNGA-like     &&& SAMI-like                                                   \\
\hline
& \multicolumn{5}{c}{Hot, fast rotating corona}                                              \\
\hline
\textit{Planck}     && 6.9e4 | 2.2e4  &&& 5.9e4 | 2.7e4  \\
ACT                 && 4.4e2 | 3.5e2  &&& 1.6e3 | 1.0e3  \\
Simons              && 2.8e2 | 6.7e1  &&& 1.1e3 | 2.2e2  \\
CMB-S4              && 2.5e2 | 4.0e0  &&& 9.4e2 | 1.4e1  \\
CMB-HD              && 6.0e0 | 1.0e0  &&& 1.2e1 | 2.0e0  \\    
\hline
& \multicolumn{5}{c}{multi-phase, slow rotating corona}  \\
\hline
\textit{Planck}     && 5.3e6 | 1.6e6  &&& \emph{7.5e6} | 3.5e6  \\
ACT                 && 4.9e4 | 3.1e4  &&&       2.3e5  | 1.4e5  \\
Simons              && 3.1e4 | 6.1e3  &&&       1.4e5  | 2.9e4  \\
CMB-S4              && 2.7e4 | 4.1e2  &&&       1.3e5  | 2.0e3  \\
CMB-HD              && 1.6e3 | 7.8e1  &&&       7.1e3  | 4.0e2  \\
\hline
\hline
\end{tabular}
\caption{The number of galaxies required for a $3\sigma$ detection of a rkSZ signal. For each combination of CMB experiment and galaxy survey type, the number on the left corresponds to the aperture filter (measured at 0.1\,$R_{\rm vir}$) and the one on the right to matched filtering. The combination of {\it Planck} and a SAMI-like survey, using the aperture filter and adopting the
multi-phase slow-rotator model, yields a required number of galaxies exceeding those available within the stellar-mass and redshift range of the galaxy survey. This detection is therefore impossible, and is marked in italics (see \S~\ref{rksz1.discussion.feasability} for discussion).
}
\label{tab:rksz1.forecast.results}
\end{table}

\section{Stacking \textit{Planck} data at the positions of MaNGA galaxies}\label{rksz1.mangaplanck}
As a proof-of-concept, we stacked \textit{Planck} data at the positions of galaxies from the MaNGA survey. While the number of galaxies is insufficient to make a rkSZ detection, it can yield an upper limit on the average CMB temperature dipole aligned with galaxies' spin.

\subsection{Galaxy data: MaNGA}\label{rksz1.mangaplanck.manga} 
MaNGA is an integral field survey with the goal to acquire spatially-resolved spectroscopy from $\approx10,000$ galaxies \citep{Bundy2015}. Galaxies were targeted to follow a (roughly) flat distribution with respect to their stellar mass, along two different sequences \citep{Wake2017}. The first, or ``Primary" sample, consists of low-redshift galaxies for which MaNGA's IFU spectrographs cover $\approx 1.5$ times their half-light radius, $R_e$. The second, or ``Secondary" sample, is comprised of higher-redshift galaxies, for which MaNGA's spatial coverage increases to $\approx 2.5$\,$R_e$. A third sample, the ``Color-enhanced supplement", increases the survey's galaxy count in areas of the Primary sample's color-magnitude diagram that are otherwise poorly sampled. Galaxies in the Primary sample contribute the most to the overall SNR of the stacked kSZ signal, because its galaxies have larger apparent size on the sky (for a given stellar mass), and therefore their kSZ signal is suppressed less by the CMB beam's smoothing.

Among other data products, MaNGA provides, for each observed galaxy, two-dimensional maps of line-of-sight velocities separately for stars and gas. This kinematic information can be used to estimate the galaxy's spin angle projected on the sky, hence the usefulness of this survey to try to detect any effect (on average) of the galaxies' rotation.

After applying a series of quality cuts described in Appendix~\ref{rksz1.appendix.manga} to MaNGA's {\tt DRPall} catalog, made publicly available as part of SDSS's data release DR15, our stacking sample consists of 2,664 galaxies: 1,231 are part of the Primary sample, 982 are part of the Secondary sample, and 451 of the Color-enhanced supplement.

We used additional value-added catalogs to access information about the galaxies in our stacking sample that is not included in the {\tt DRPall} catalog. The MaNGA Morphology Deep Learning DR15 catalog supplies information on the galaxies' morphology. The morphological classification is performed using an automated model trained and tested on SDSS-DR7 images \citep{Dominguez2018}. The 2,664 galaxies are split by type as follows: 282 are ellipticals, 424 are S0s, 1,953 are spirals, and 5 are classified as irregulars. 

Finally, the Galaxy Environment for MaNGA Value Added Catalog ({\tt GEMA-VAC}, \citep{Argudo2015}) provides environmental information based on the sign of the eigenvalues of the tidal tensor at the location of each galaxy \citep{Hahn2007}. In our stacking sample, 1,056 galaxies are in a cluster environment, 1,239 in filaments, 331 in sheets, and 38 in voids.

\subsection{CMB data: \textit{Planck}}\label{rksz1.mangaplanck.planck}

We used publicly available CMB maps from \textit{Planck} \citep{Planck2018}, specifically the full mission, temperature {\tt SMICA-noSZ} map and full mission, single-frequency maps from \emph{Planck}'s high frequency instrument. The {\tt SMICA-noSZ} map is a linear combination of multi-frequency CMB maps that cancels any contribution with the spectrum of the tSZ effect, leaving the kSZ signal, which has the same spectrum as the CMB, unaffected. It also cleans other foreground signals, based on their contribution to the variance in the data.

On average, the tSZ signal from galaxy halos should be isotropic. While measurements with the aperture filter described in \S~\ref{rksz1.stats.aperture} are insensitive to isotropic signals, measurements based on a matched filtering could be affected, as we discuss in \S~\ref{rksz1.discussion.modellimitations}. It is therefore preferable to use data that has already been cleaned from other halo-induced temperature anisotropies, such as the tSZ. The {\tt SMICA-noSZ} map has a HEALPix resolution $N_{pix}=2048$ and a spatial angular resolution of 5 arcmin FWHM. We combined the temperature data with the common temperature confidence mask before performing any measurement, and verified that our results do not depend on the mask used.

\subsection{Stacking}\label{rksz1.mangaplanck.stack}
CMB data needs to be aligned with the galaxies' projected spin angle. Otherwise, any possible rkSZ signal would be cancelled. We also scaled the \textit{Planck} data with the angular diameter of each galaxy's halo, to add the signal profiles coherently.

Since we are interested in capturing the rotation of ionized gaseous halos, we estimated the spin angle using the emission line with the shortest wavelength (highest photon energy) for which MaNGA provides kinematic information: O\,II. For both of the two lines that form the doublet (3,727\,\r{A} and 3,729\,\r{A}, corresponding to temperatures of $3.8\times10^4$\,K), we computed the momentum of their line-of-sight velocity map relative to the galaxy's position in the catalog:

\begin{equation}\label{eq:rksz1.mangaplanck.stack.momentum}
    \mathbf{L}=\sum_i F_i \mathbf{v}_i \times \mathbf{r}_i
\end{equation}

For each spaxel, $i$, the line-of-sight velocity, $\mathbf{v}_i$, and the flux, $F_i$, are weighted with their inverse variance before entering the cross-product with the spaxel's position vector, $\mathbf{r}_i$. The projected spin angle for each line is the angle between $\mathbf{L}$ and the North direction. The spin angle for the galaxy is the average of the spin angle for each of the doublet's lines.

We confirmed that our results do not depend on the specific line used to estimate the spin angle. We reached the same results using Ne\,III, and angles estimated from all emission lines measured by MaNGA are highly correlated with each other. This is not surprising, since MaNGA probes the inner regions of halos (up to $\approx 1.5 \, R_e$ or $\approx 2.5 \, R_e$), where gas kinematics tend to be coherent. One of the criteria used to select the galaxies used for this analysis was precisely that the spin angle did not depend strongly on the tracer used to compute it; we rejected galaxies for which the spin angle estimated from O\,II differed significantly from the one estimated from H$_{\alpha}$ (see Appendix~\ref{rksz1.appendix.manga}). A visual inspection showed that most of these rejected galaxies have complex kinematics or are face-on systems that would not contribute to the rkSZ signal.

MaNGA measures the gas kinematics only in the inner regions of galaxies, on average up to $\approx2\%$ of the galaxies' virial radius ($\approx5-6\%$ in the most highly resolved cases). By comparison, the rkSZ signal peaks further out, in the inner halo, at $0.1-0.2$\,$R_{\rm vir}$ (depending on the CMB experiment's beam size). This corresponds to an extrapolation by a factor of several in spatial scale; a key assumption in our analysis is therefore that the galaxies' gaseous halos at this radius co-rotates with their inner region where the spin is measured. In a hierarchical formation scenario, this is not necessarily the case, for outer halos may have built up from contributions with different angular momenta \citep{Stewart2017}. Nevertheless, recent measurements on simulations show a strong correlation in the angular momentum of hot gas across most of the virial radius ~\citep{DeFelippis2020}. A non-detection can then be due to the lack of correlation between the rotations of the inner and outer regions of gaseous halos. However, in order to significantly suppress the rkSZ signal, the r.m.s. variation in the estimated spin angle would have to be large (see \S~\ref{rksz1.discussion.obs_uncertainties} and Appendix~\ref{rksz1.appendix.filtermisplacement}).

We scaled the CMB data with the angular diameter of each galaxy's host DM halo. This angle is fully determined by the galaxies' redshifts and their stellar masses, both found in MaNGA's {\tt DRPall} catalog. Stellar masses, which are inferred from the galaxies' Sersic photometry (hence their $h^{-2}$ cosmological dependency), were converted to DM halo masses and virial radii (see \S~\ref{rksz1.model}).

Matched filtering requires a template for the expected signal. We built two templates for each galaxy, based on the hot fast-rotator and multi-phase slow-rotator models described in \S~\ref{rksz1.model}. A parameter that needs to be derived from the data to generate the templates is the galaxies' inclination angle, $i$. To do so, we modeled galaxies as oblate spheroids, for which \citep{Holmberg1958}
\begin{equation}\label{eq:rksz1.mangaplanck.stack.inclination}
    \cos i = \sqrt{\frac{\frac{b}{a}-q^2}{1-q^2}}.
\end{equation}
We used the Sersic axial ratio, $b/a$, from the {\tt DRPall} catalog, and assumed that galaxies have aspect ratios of $q=0.15$ when seen edge-on. Furthermore, we assigned an inclination of 90$^\circ$ to all galaxies whose edge-on probability reported in the morphology catalog exceeds 99\%. This estimation is meaningless for elliptical galaxies.

Another reason why elliptical galaxies are problematic for our stacking analysis is that they have, in general, small spin parameters. For these reasons, in addition to the stack analysis of the 2,664 galaxies selected from MaNGA's {\tt DRPall} catalog, we also performed an analysis restricted to the 1,953 spiral galaxies in the sample. Furthermore, the environment can affect the properties of galactic gaseous halos. The kinematics of the outer halos of spiral galaxies in clusters may be perturbed by close encounters with other cluster members and interactions with the intra-cluster medium. Thus, we performed a third analysis restricted to spiral galaxies which do not reside in a cluster environment (field spirals, 1,235 galaxies).

\begin{figure*}
    \centering
    \includegraphics[width=\linewidth]{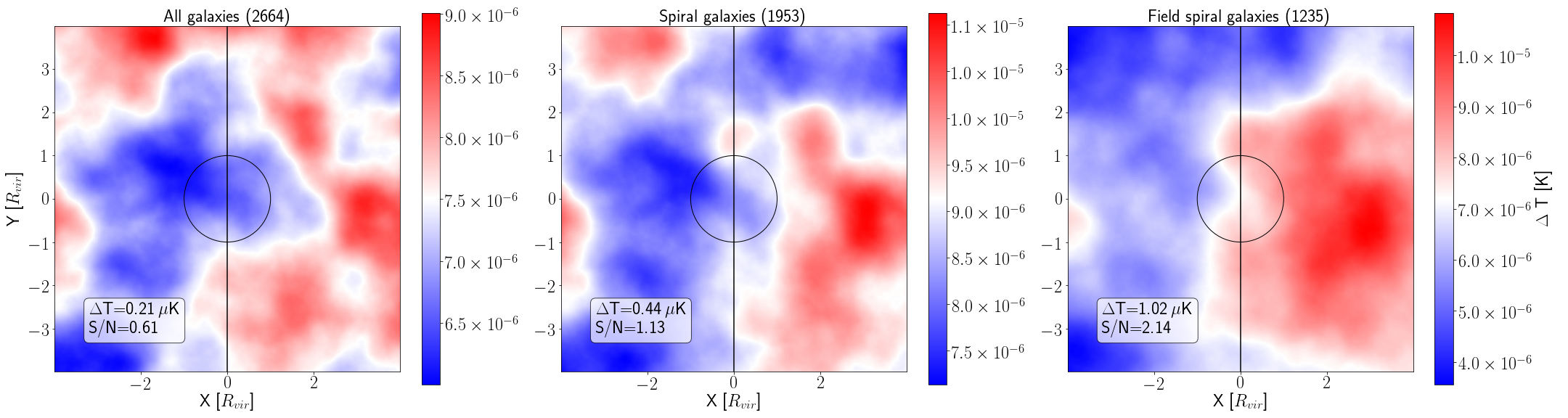}
    \caption{
    \textit{Planck} {\tt SMICA-noSZ} CMB data stacked on the positions of MaNGA galaxies (with equal weight) after aligning with the galaxies' spin angles, and scaling to their $R_{\rm vir}$. The left panel shows the results for all 2,664 galaxies in our sample, the central panel for the 1,953 spiral galaxies, and the right panel for the 1,235 field spiral galaxies.
    For each galaxy, located at the origin (0,0), we stacked spin-aligned $8\times8\,R_{\rm vir}$ patches. The circles correspond to $R_{\rm vir}$ and the vertical lines mark the expected galaxy spin direction (pointing towards the~-y axis, the right half is approaching the observer, and the left side receding). Any rotation-induced temperature dipole should show a left-right cold-hot temperature asymmetry (see Fig.~\ref{fig:rksz1.model.signal}).
    In each text box, the measured dipole on a $1.0\,R_{\rm vir}$ aperture, and its signal-to-noise ratio based on the theoretical noise calculation described in ~\ref{rksz1.stats.aperture}.
    }
    \label{fig:rksz1.mangaplanck.stack.aperture}
\end{figure*}

The results of stacking \textit{Planck}'s CMB data on the positions of these three galaxy samples are displayed in Fig.~\ref{fig:rksz1.mangaplanck.stack.aperture}, for the case in which all galaxies are equally weighted. The results of the analyses do not change when optimal weights based on the expected signal are used. Visually, the stack corresponding to field spirals (right panel) shows a temperature asymmetry with the correct sign if it were induced by the rotation of the galaxies (the spin angle points towards the negative Y axis). Its amplitude ---$1.02\,\mu$K measured on a filter with an aperture of $1.0\,R_{\rm vir}$--- is large compared with the expected signal ---a mere $2.7\times10^{-2}\,\mu$K if the hot, fast rotator is a good model---, but still only $2.14\times$ larger than the expected noise for that aperture. The other two samples (middle and left panels) appear consistent with noise.

\begin{figure}
    \centering
    \includegraphics[width=\linewidth]{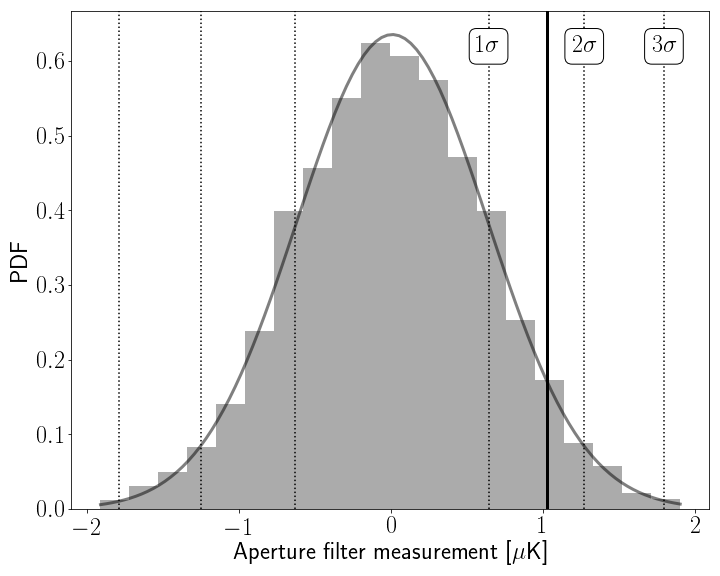}
    \caption{
    Probability distribution function (PDF) of the temperature dipole of the {\tt SMICA-noSZ} map stacked on the positions of 1,253 field spirals after randomizing their spin angles. The PDF is inferred from $10^4$ measurements using $1.0\,R_{\rm vir}$ aperture filters that have been randomly placed and rotated. Superimposed, a Gaussian PDF with the same mean and standard deviation as the $10^4$ measurements, shows good agreement with the data. Dotted vertical lines correspond to 1$\sigma$, 2$\sigma$, and 3$\sigma$ thresholds. The solid vertical line is the measured dipole on the stack with the galaxies' spins aligned.
    }
    \label{fig:rksz1.mangaplanck.stack.aperture_bootstrap}
\end{figure}

\begin{figure*}
    \centering
    \includegraphics[width=\linewidth]{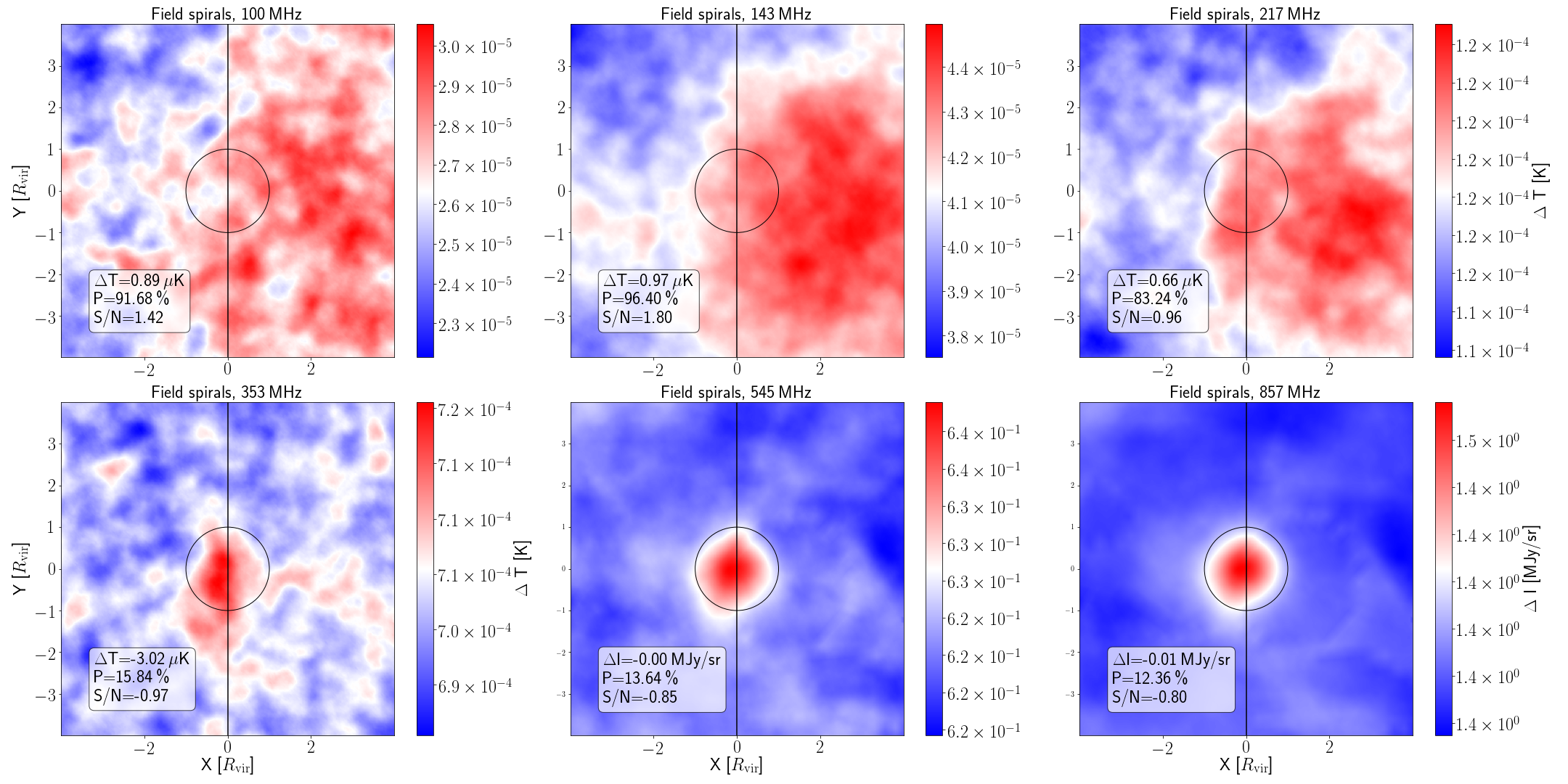}
    \caption{
    Single-frequency \textit{Planck} CMB maps from its high-frequency instrument (HFI), stacked with equal weights on the positions of the 1,235 field spirals after aligning their spin angles. Note the difference in units for the two highest-frequency maps: MJy/sr instead of K. As with Fig.~\ref{fig:rksz1.mangaplanck.stack.aperture}, the circle represents the average virial radius and the vertical line the spin direction. The text boxes show the measured dipole, as well as its percentile and S/N ratio. The percentile and significance have been derived from noise-only stacks.
    }
    \label{fig:rksz1.mangaplanck.stack.singlefreq}
\end{figure*}

An alternative way to measure the significance of any dipole measured on stacked data is to draw a large number of measurements on noise-only data, after randomizing the positions and orientations of the filter. We show the result for $10^4$ such measurements in Fig.~\ref{fig:rksz1.mangaplanck.stack.aperture_bootstrap}. The noise-only map on which the measurements were done is the result of stacking the CMB data on the positions of the galaxies assuming a random orientation of their spins. The measurements follow a Gaussian distribution, and the measured dipole of $1.02\,\mu$K corresponds to a 94.65 percentile, or a significance of 1.63$\sigma$. This is lower than the $2.14\sigma$ significance derived from the theoretical calculation of the noise for that aperture. While not significant enough to be a detection, the measured dipole is suggestive enough to raise the question of whether it is real, and if so, what could cause such an unexpectedly large signal. To answer that question, we performed the same analysis on \textit{Planck}'s single frequency maps (see Fig.~\ref{fig:rksz1.mangaplanck.stack.singlefreq}).

In the 100, 143, and 217 MHz data, the stacks look similar to the one from the {\tt SMICA-noSZ} map. The significance of the measured dipoles, estimated from $10^4$ measurements on noise-only stacks (stacks with randomized galaxy spins), is lower than that of the {\tt SMICA-noSZ} dipole, at 1.42$\sigma$, 1.80$\sigma$, and 0.96$\sigma$, respectively. Higher frequency maps, at 353, 545, and 857 MHz, look different, with a clear signal coming from within the galaxies' virial radius, and a small dipole with of opposite sign to the one measured on the {\tt SMICA-noSZ} map. The lack of a consistent dipole across frequencies, and the low significance of the measurements indicate that there is no real temperature dipole in our data above the noise level. These results are robust to the choice of CMB mask, size of the CMB patches and their weighting for stacking. Finally, matched filtering, using the two models for galactic atmospheres described as part of this work, does not uncover a signal either.

\section{Discussion} \label{rksz1.discussion}

\subsection{Model  and observational uncertainties} \label{rksz1.discussion.modellimitations}
The estimates shown in Table~\ref{tab:rksz1.forecast.results} do not include model uncertainties. The models used to describe the gaseous component of galactic halos are a simplification. For instance, the density profiles are spherically symmetric and the pressure profiles do not account for any rotational support. For a feasibility study, and given the current uncertainties about the properties of galactic atmospheres, we deem the level of detail of the kSZ models sufficient, and leave for future work the use of more sophisticated models, such as the self-consistent rotating profiles developed in \citep{Sormani18}, or models taken directly from hydrodynamical simulations.

Two model parameters that could modify significantly the strength of the kSZ signal are the total mass of baryons in the halo, parametrized by the halo baryon fraction, and the galaxy's peculiar velocity. An effect on halo scales that we have not included in our models is the tSZ.

Our models assume that the halo baryon abundance matches the cosmic average, $M_b = f_b M_h \equiv (\Omega_b/\Omega_m) M_h$. In practice, only a small fraction of these baryons are observed in galaxies, which is known as the missing baryon problem. These baryons could reside in the IGM, and the galactic halos could, as a result, be baryon-poor relative to the cosmic abundance. A lower baryon fraction would reduce the electron density in the halo, and therefore its rkSZ signal. 

For the limiting case in which all the gas is ionized, the electron density scales linearly with the baryon fraction through the multiplicative factor of $\rho_0$ in Eq.~\eqref{eq:rksz1.model.ne.hotgas}. A baryon fraction of half the cosmic abundance would reduce electron densities and the kSZ signal by a factor of two and the number of galaxies needed for a detection would increase by a factor of four. For the more realistic, multi-phase model, the impact of the overall baryon fraction in the halo is more complex. The cooling density in Eq.~\eqref{eq:rksz1.model.ne.cooling} is independent of the total baryonic mass of the halo (as long as DM dominates the gravitational potential and determines the temperature of the hot gas in hydrostatic equilibrium). The cooling radius depends on the baryon fraction, since it is partly determined by the initial distribution of hot gas, and it affects the density profile  of the hot corona through $\xi_c$ in Eq.~\eqref{eq:rksz1.model.ne.hotcorona}. As a result, the residual hot corona in the multi-phase model is less sensitive to the baryon fraction than the gas in the hot model. Physically, in the multi-phase model, the reduction in the baryons is primarily absorbed by the cold gas, except in the outer regions, where the small densities contribute little to the SNR of the rkSZ signal.

\begin{figure}
    \centering
    \includegraphics[width=\linewidth]{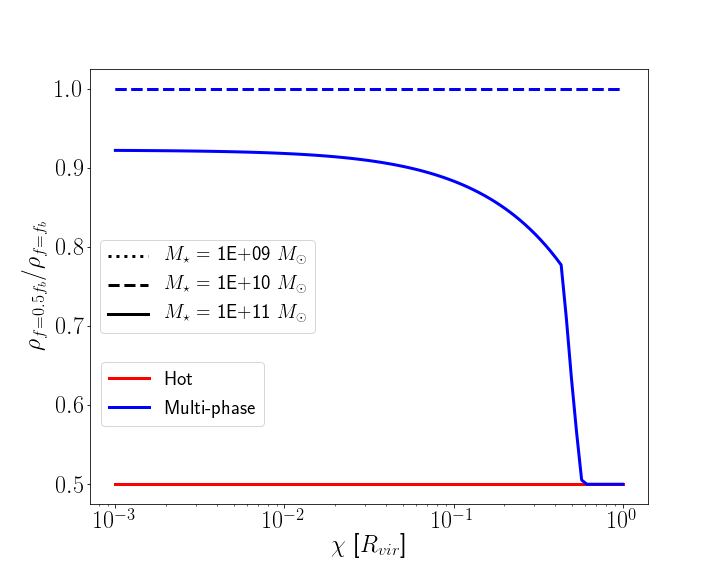}
    \caption{The figure illustrates the impact of a factor-of-two reduction in the total baryon mass in the halo relative to the cosmic mean value.  The corresponding reduction in the density of the hot, ionized gas component is shown as a function of radius, in units of the virial radius. Results are shown for the same three halos as in the previous figures, labelled by the stellar mass of their central galaxies. In the hot model, where all the baryons are ionized, the gas density scale linearly with the baryon fraction, independent of mass and radius (shown in red). The hot coronae of the multi-phase gaseous halos (shown in blue) are less affected, except in the outer regions where the rkSZ signal contributes little to the SNR. The effect on the multi-phase model decreases with smaller halo mass.
    }
    \label{fig:rksz1.discussion.modellimitations.baryonfraction}
\end{figure}

Fig.~\ref{fig:rksz1.discussion.modellimitations.baryonfraction} shows the effect of a reduced baryon fraction on the free electron density. A reduction in the baryon fraction by a factor of two reduces the electron density in the inner parts of multi-phase halos only by $\lsim 10\%$. Consequently, the number of galaxies needed for a detection may not be as sensitive to the baryon fraction in galactic halos may naively be expected. 

When estimating the number of galaxies needed for a detection, we also did not take into consideration the galaxies' peculiar velocities. Their line-of-sight velocity, relative to the Hubble flow, can be significant \citep{Strauss1995} and the induced kSZ effect dominant, compared to the rkSZ. Nevertheless, since the induced temperature shift is symmetric with respect to the galaxies' spin axes, the aperture filter is blind to this shift. In the absence of centering errors, matched filters are also insensitive to a symmetric signal. The reason is that the matched filter has odd parity relative to the spin axis, and therefore a convolution with a signal that has even parity relative to the same direction leaves it unchanged (including the height of the central peak). In the presence of centering errors (or anisotropies of the kSZ signal induced by the peculiar motions), this conclusion no longer holds. Fortunately, when stacking a large ensemble of galaxies with random peculiar velocities, the effect should still average out. 

The tSZ signal from galactic halos is symmetric relative to their spin axes, like the kSZ from peculiar motions, but contrary to the peculiar-motion-induced kSZ, its sign is not random. Subsequently, in the presence of filter centering errors, its effect on matched filtering will not average out, and measurements will be affected. The aperture filter is still insensitive to this effect. When applying matched filtering to CMB data for stacking, either the tSZ signal should be modelled and incorporated in the analysis, or the CMB data should be cleaned from the tSZ signal. This is the reason we used the {\tt SMICA-noSZ} map in \S~\ref{rksz1.mangaplanck}.

The significance of any future detection should be assessed by considering these factors, together with any observational uncertainties in the parameters that inform the models, such as spin angles, stellar masses, scatter in the stellar mass to halo mass relation, etc. Such an analysis is beyond the scope of the present study but will be warranted if/when a detection is claimed.

\subsection{Measurement uncertainties} \label{rksz1.discussion.obs_uncertainties}
Misplacing the filters used to measure the rkSZ effect relative to the signal is a source of errors, independently of the model used to interpret the measurements. A filter misplacement can be a decentering relative to the galaxy, a misalignment of the filter's axis with the galaxy's projected spin vector, or a combination of the two. The average effect of these misplacements (assuming they are random) can be characterized as a suppression of the measured kSZ signal. If the probability distribution of these positioning and alignment errors, $\mathbf{\epsilon}$, is given by $p(\mathbf{\epsilon})$, the expectated value of the dipole measurement is
\begin{equation}\label{eq:rksz1.stats.aperture.meansignal}
    \langle s \rangle_{\mathbf{\epsilon}} = \int_{-\infty}^{\infty} \textrm{d}\mathbf{\epsilon} \,\, p(\mathbf{\epsilon}) s(\mathbf{\epsilon}),
\end{equation}
where $s(\epsilon$) is the signal computed from Eq.~\ref{eq:rksz1.stats.aperture.dipole} in the presence of an uncorrected error $\epsilon$.
In order to assess the sensitivity of the aperture dipole statistic to these errors, we assume that position offsets and misalignment angles are both normally distributed, fully specified by their  full-width-at-half-maximum (FWHM), and zero mean (i.e. no systematic spatial offsets or misalignments). 

The sensitivity of the filters to decentering errors depends significantly on the resolution of the CMB experiment, as detailed in Appendix~\ref{rksz1.appendix.filtermisplacement}. Positioning errors with FWHM=$0.2\,R_{\rm vir}$ ($\sim 1\,$arcmin), hardly suppress the signal for experiments that barely resolve galactic halos (such as \textit{Planck}), and can yield a reduction of up to $\sim10$\% for ACT. The aperture filter can be more sensitive than a matched filter, for high-resolution experiments, with a signal suppression of up to 40\% for such a decentering on a CMB-S4 experiment.

The robustness of the aperture and matched filters to misalignments with respect to the true galaxies' atmospheres' spin angle does not depend strongly on the experimental configuration. Large errors in the spin angle with FWHM=$90\,$deg result in a signal suppression of less than 25\%.

\subsection{Detection feasibility} \label{rksz1.discussion.feasability}
The number of galaxies required for a 3$\sigma$ detection shown in Table~\ref{tab:rksz1.forecast.results} raises the question of how feasible is the measurement of a large number of galactic spins, and whether the rkSZ signal could be masked by other effects.

Table~\ref{tab:rksz1.discussion.feasability.footprint} shows the same information as Table~\ref{tab:rksz1.forecast.results}, with number of galaxies converted to an equivalent survey sky coverage. To do so, we use the double Schechter local galaxy stellar mass function ~\citep{Baldry2008}:

\begin{equation}\label{eq:rksz1.discussion.feasability.smf}
    n = \phi_1 \Gamma \left(\alpha_1 + 1, \frac{M_1}{M^{\star}}, \frac{M_2}{M^{\star}}\right) + \phi_2 \Gamma \left(\alpha_2 + 1, \frac{M_1}{M^{\star}}, \frac{M_2}{M^{\star}}\right),
\end{equation}
where $\Gamma$ is the incomplete gamma function,
\begin{equation}\label{eq:rksz1.discussion.feasability.gamma}
    \Gamma(x, a, b) = \int_a^b \mathrm{d}t \exp(-t) t^{x-1}.
\end{equation}
and the fitted values for the parameters are: $\phi_1=4.26\times10^{-3}\textrm{Mpc}^{-3}, \alpha_1=-0.46, \phi_2=0.58\times10^{-3}\textrm{Mpc}^{-3}, \alpha_2=-1.58$ and $M^{\star}=10^{10.648} M_{\odot}$. 

For each of the survey bins shown in Figure~\ref{fig:rksz1.forecast.surveys}, we multiply this galaxy number density by the corresponding comoving survey volume, taking into consideration the full sky solid angle ($4\pi\,$sr). The resulting number of available galaxies is $5.76\times10^6$ for a MaNGA-like survey and $6.58\times10^6$ for a SAMI-like survey. To have an idea of the depth required, the faintest galaxy in the MaNGA Primary sample has a magnitude of $G$=18.68. Upcoming CMB experiments should be capable of detecting the rkSZ signal with complete surveys of low-redshift galaxies covering a few thousands of deg$^2$.

\begin{table}
\centering
\begin{tabular}{l r r c c r}
           && MaNGA-like     &&& SAMI-like  \\
\hline
& \multicolumn{5}{c}{Hot, fast-rotating corona}   \\
\hline
\textit{Planck}     && 497 | 154  &&& 369 | 168  \\
ACT                 &&   3 |   3  &&&  10 |   6  \\
Simons              &&   2 |$<$1  &&&   7 |   1  \\
CMB-S4              &&   2 |$<$1  &&&   6 |$<$1  \\
CMB-HD              &&$<$1 |$<$1  &&&$<$1 |$<$1  \\
\hline
& \multicolumn{5}{c}{multi-phase, slow-rotating corona}             \\
\hline
Planck              && 38,239 | 11,759 &&&     X | 21,717  \\
ACT                 &&    352 |    222 &&& 1,449 |    869  \\
Simons              &&    219 |     44 &&&   896 |    184  \\
CMB-S4              &&    194 |      3 &&&   789 |     12  \\
CMB-HD              &&     11 |      1 &&&    45 |      3  \\
\hline
\hline
\end{tabular}
\caption{
Minimum footprint size of surveys required for a $3\sigma$ detection of a rkSZ signal, in deg$^2$.  This corresponds to the required number of galaxies, shown in Table~\ref{tab:rksz1.forecast.results}, and converted to sky coverage using the stellar mass function described in \S~\ref{rksz1.discussion.feasability}. An "X" indicates that the required sky area exceeds the full sky. The number on the left corresponds to the requirement using an aperture filter (measured at 0.1\,$R_{\rm vir}$) and the one on the right to the use of matched filtering.}
\label{tab:rksz1.discussion.feasability.footprint}
\end{table}

In the absence of spectra, the orientation of the projected spin parameter could, in principle, still be estimated from photometry. From MaNGA data, the position angle of the single-component Sersic fit in the $r$-band is highly correlated with the projected spin angle we estimated using the O\,II emission line. We find that the standard deviation of the difference between the photometric position angle and the O\,II derived angle is just 9.0$^\circ$. However, the morphological direction from photometry alone leaves the sense of the rotation, which is crucial for attempts to measure the rkSZ signal, undetermined. On the other hand, the spectroscopic requirement to discern the sense of the projected spin, given its orientation, will be much less demanding than the ones reached in recent IFU surveys. Lower-resolution, faster surveys could be envisioned, maybe even narrow-band imaging, to detect the Doppler asymmetry on the two sides of a galaxy whose orientation is known. 

Future H\,I surveys will provide additional information on gas kinematics for a large number of nearby galaxies, extending beyond their stellar component. For example, the WALLABY survey (Widefield ASKAP L-band Legacy All-sky Blind SurveY, \footnote{Widefield ASKAP L-band Legacy All-sky Blind SurveY:\url{https:https://www.atnf.csiro.au/research/WALLABY/}}) will cover $\approx 31,000$\,deg$^2$, up to a redshift of $z<0.26$, and could marginally resolve $\approx 540,000$ galaxies \citep{Duffy2012}. We expect that such marginally resolved observations could be used in isolation or combined with photometric surveys to break the spin-orientation uncertainty. Existing H\,I data has already been used to estimate galactic spins, showing that they are aligned with cosmic filaments \citep{Blue2019}.

Even with a number of galaxies' spin angle measurements sufficient to beat down the instrumental noise and the primary CMB anisotropies, alternative signals may mask that from the rkSZ effect. Two possibilities are the Birkinshaw-Gull effect (BG) and thermal dust
emission.

The BG effect ~\citep{Birkinshaw1983, Hotinli19, Yasini2019, Hagala2019} induces a temperature dipole, of a magnitude comparable to that of the rkSZ effect, with a decrease in CMB temperature following the galaxies' transverse proper motion (and a temperature increase opposite to the transverse proper motion). Regardless of the presence of systematic alignments between the directions of galaxies' spins and their proper motions ---induced, for instance, by filaments \citep{Tempel2013, Krolewski2019, Welker2020, Blue2019}---, we don't expect any systematic alignment in their sense. Thus, on spin-aligned stacks of large samples of galaxies, any BG-induced temperature dipole should average to zero. 

Another possible source of contamination is Doppler shifted thermal emission from dust co-rotating with the galaxy. There is evidence for the presence of dust in galactic halos ~\citep{Menard2010} and it could co-rotate with the galaxies. Such a contaminant can be separated, in principle, using multi-frequency CMB measurements. In the worst-case scenario, dust emission could be used on its own as a tracer of galactic rotation in those regions within galactic halos where dust is kinematically coupled to gas. In the \textit{Planck}'s high frequency stacks (see lower panels of Fig.~\ref{fig:rksz1.mangaplanck.stack.singlefreq}), we see what appears to be an (unresolved) signal which would combine dust emission and tSZ effect from the field spiral galaxies (see a similar detection at galaxy cluster level in \citep{Melin2018}). There is no clear indication of a Doppler-shifted dipole in these maps either. 

The non-detection of a significant temperature dipole around field spirals using \textit{Planck} data, puts a 3$\sigma$ upper limit on the average temperature dipole around field spirals of $1.9\,\mu$K (measured on a $1\,R_{\rm vir}$ aperture). In contrast, several studies have measured CMB temperature anisotropies aligned with the rotation axis of nearby spiral galaxies of a few tens of $\mu$K \citep{DePaolis2014, DePaolis2016, Gurzadyan2015, Gurzadyan2018}. This discrepancy may be due to the shape of the anisotropy (a highly concentrated temperature anisotropy can be significantly suppressed when unresolved by the detector's beam), or may indicate that the nearby spirals with (large) measured temperature anisotropies are not representative of the MaNGA sample used for our analysis. The magnitude of the CMB temperature asymmetry measured around nearby galaxies is too large to be caused by the rkSZ, and we refer to \citep{DePaolis2014, DePaolis2016, Gurzadyan2015, Gurzadyan2018} for a brief enumeration of possible causes.

\section{Conclusions} \label{rksz1.conclusion}

In this paper, we analyzed the feasibility of detecting the kSZ signal from the coherent rotation of the gaseous halos of galaxies.  Such a detection would provide novel insight into the angular momentum distribution of gas, and more broadly, into galaxy formation.  Our analysis is based on two models for galactic atmospheres: a fully ionized gaseous halo rotating at the DM halo's circular velocity, which can be considered an upper bound, and a more realistic model consisting of a multi-phase gaseous halo rotating at a fraction of the DM halo's circular velocity. As a proof-of-concept, we stacked public \textit{Planck} CMB data on the positions of $\approx2,000$ MaNGA galaxies after aligning these data with the galaxies' locations and spins, and scaling them to their halos' expected angular diameters.  Our main findings can be summarized as follows:

\begin{itemize}
    \item The number of galaxies required for a $3\sigma$ detection with current CMB data is large, at best $2.2\times10^4$, beyond the largest set of galactic spin measurements currently available.  The primary limitation is angular resolution in the CMB data, which only marginally resolves gaseous halos. 
    \item Upcoming  high-resolution, low-noise CMB experiments will significantly reduce the required number of galaxies. A galaxy survey measuring the spins of nearby galaxies covering $\sim10\,$deg$^2$ could be sufficient to rule out upper-bound models, and a few hundreds of deg$^2$ should be sufficient to detect galaxies' rkSZ effect.
    
    \item The use of matched filtering can reduce the number of galaxies needed up to an order of magnitude for future CMB experiments. Such measurements can be sensitive to signal modeling, particularly in the presence of non-random CMB temperature anisotropies at the scales of galactic halos, such as those induced by their thermal SZ effect.
    
    \item A stacking analysis of \textit{Planck} CMB data on the position of MaNGA galaxies rules out average non-random temperature dipoles aligned with the spin angles of field spirals down to 1.9\,$\mu$K. This may be inconsistent with asymmetries of up to $\approx100$\,$\mu$K measured in nearby spiral galaxies (e.g. M31). 
    
    \item If Doppler shifted, anomalous thermal dust emission is responsible for the measured asymmetries in nearby spirals, as is claimed in a recent study \citep{Amekhyan2019}, it could mask the rkSZ signal induced by gaseous galactic halos. This contaminant could be removed before searching for the kSZ signal. Alternatively, the emission from dust itself could be used to trace the kinematics of hot gas.
    
\end{itemize}

We conclude that the rotational kinematic Sunyaev-Zeldovich signal, imprinted on the cosmic microwave background by spinning hot gas in galactic halos, is a promising and novel probe of galaxy formation, and should be feasible to detect in future, high-resolution CMB surveys, combined with estimates for the spin orientations of $\lsim 10^4$ galaxies.   A much larger number of galaxies would allow studies of the dependence of the angular momentum of the gas on galaxy properties.

\section{Acknowledgements} \label{ksz1.acknowledgements}
We are grateful to the anonymous referee for detecting a numerical mistake in a previous version of this manuscript, and the insightful comments provided. We thank Greg Bryan, Daniel DeFelippis, Colin Hill and Mary Putman for their suggestions and discussions, and Siavash Yasini brought to our attention the possible role of the BG effect as a contaminant to measurements of the rkSZ. We gratefully acknowledge support from NASA ATP grant 80NSSC18K1093, and the use of the NSF XSEDE facility for simulations and data analysis in this study.

\bibliography{Rotation_kSZ}

\clearpage
\newpage

\appendix 

\section{Impact of filter misalignment and centering errors.}
\label{rksz1.appendix.filtermisplacement}

Even in the absence of noise, the true temperature dipole induced by the rkSZ effect can differ from the one measured for any given galaxy. A centering error in the aperture filter described in \S~\ref{rksz1.stats.aperture}, and/or a misalignment between the filter's axis and the galaxy's projected spin vector, will suppress the measured dipole. The same applies when convolving a matched filter with the CMB data. While for a single galaxy the maximum response to the matched filter localizes the center of the galaxy's halo, when stacking the data for many noise-dominated galaxies, their center needs to be chosen a priori.

\begin{figure*}
    \centering
    \includegraphics[width=\linewidth]{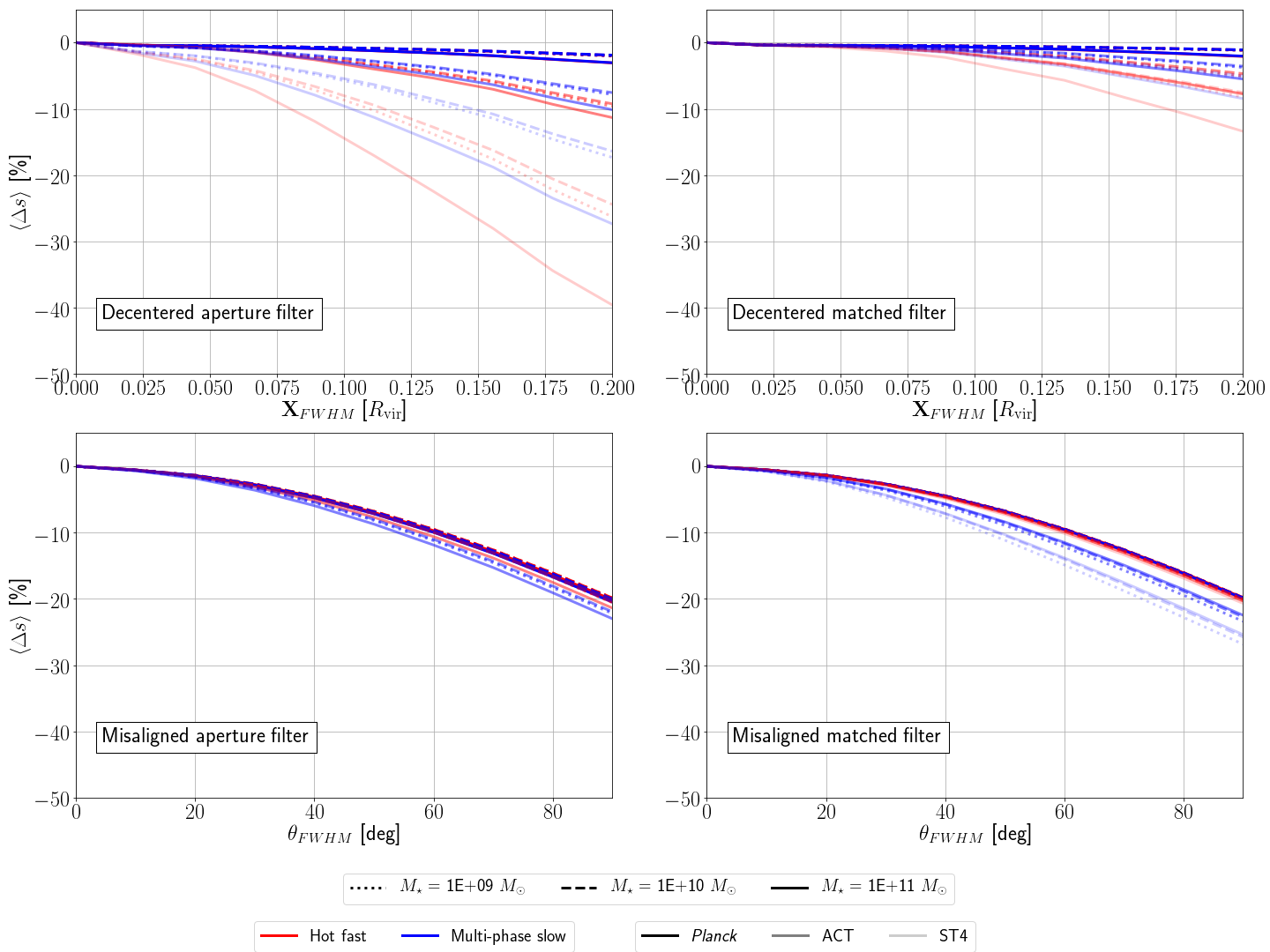}
    \caption{
    Mean suppression in the measured rkSZ signal due to decentering relative to the halo position and to misalignment relative to the galaxy's projected spin angle. Color indicates whether the rkSZ signal corresponds to that of a hot, fast or cold, slow rotator. The intensity of the color indicates a different CMB experiment configuration (\textit{Planck} for strong color, CMB-S4 for the faintest color and ACT for the intermediate intensity). Finally, the type of line used indicates the mass of the galaxy.\\
    \textbf{Upper panels:} Effect of a Gaussian error in the position of the filter, relative to the halo's center, as a function of the error's FWHM in units of the virial radius. On the left, effect for an aperture filter measuring the signal's dipole. On the right, effect for the measured correlation of a matched filter and the signal at the estimated (erroneous) halo center. \\
    \textbf{Lower panels:} Effect of a Gaussian error in the orientation of the filter axis relative to the galaxy's projected spin vector, as a function of the error's FWHM in degrees. As in the upper panels, on the left the effect for the aperture filter is displayed and on the right, for the matched filter.
    }
    \label{fig:rksz1.appendix.filtermisplacement.misplacement}
\end{figure*}

To assess the sensitivity of these filters to these errors, we computed the mean response of the filter according to Eq.~\ref{eq:rksz1.stats.aperture.meansignal}, assuming that the centering offsets and misalignments both follow zero-mean, normal distributions:
\begin{equation}\label{eq:rksz1.appendix.filtermisplacement.misalignment}
    \langle s \rangle_{\theta} = \frac{2}{\sqrt{2\pi}\sigma_{\theta}} \int_0^{\infty}\textrm{d}\theta \exp{\left[-\frac{\theta^2}{2\sigma_{\theta}^2}\right]} s(\theta),
\end{equation}
\begin{equation}\label{eq:rksz1.appendix.filtermisplacement.decentering}
        \langle s \rangle_{xy} =  \int_{-\infty}^{\infty}\int_{-\infty}^{\infty}\frac{\textrm{d}x \textrm{d}y}{2\pi\sigma_{xy}^2} \exp{\left[-\frac{x^2+y^2}{2\sigma_{xy}^2}\right]} s(x,y).
\end{equation}
We show in Fig.~\ref{fig:rksz1.appendix.filtermisplacement.misplacement} the mean effect on the measured signal (dipole for the aperture filter and maximum correlation for matched filter) of a filter decentering and misalignment for three galaxies of different mass, the two different atmosphere models described in S~\ref{rksz1.model}, and three different CMB experiment configurations.

The sensitivity of both filters to errors in the galaxies' spin angle estimation is similar, and does not depend strongly on the beam resolution of the CMB experiment (see lower panels of Fig.~\ref{fig:rksz1.appendix.filtermisplacement.misplacement}). Even with misalignment errors with FWHM=$90\,$deg, the signal measured by the filters will be suppressed by less than 30\% relative to its true value, regardless of the galaxies' mass, the CMB beam resolution and the galactic atmosphere model used to predict the signal.

The sensitivity to a filter decentering is strongly dependent on the resolution of the CMB experiments, the matched filter being slightly more robust than the aperture filter, in particular for high-resolution CMB experiments such as CMB-S4 ($1\,$arcmin beam). Still, for the worse case scenario, which corresponds to a high resolution experiment using an aperture filter and galaxy atmospheres that follow a hot fast rotator model, a decentering error with a FWHM=$0.2\,R_{\rm vir}$ (which corresponds to $\sim1\,$arcmin for a MaNGA-like survey, or the halo center falling outside of the galaxy) suppresses the measured signal by less than 40\%.

\section{Variance of aperture filter dipole measurements}\label{rksz1.appendix.aperturevariance}
Following \citep{Ferraro2015}, we can estimate the variance of a dipole measured over an aperture. For any given galaxy, the dipole measurement is $s=\overline{\Delta T}_R - \overline{\Delta T}_L$, where $\overline{\Delta T}_R$ is the mean temperature anisotropy measured within the right half of the aperture and $\overline{\Delta T}_L$ the same within the left half of the aperture. The CMB anisotropies have rotational symmetry, and the variance on the measured signal induced by them is given by Eqs.~\ref{eq:rksz1.stats.aperture.variance} and ~\ref{eq:rksz1.stats.aperture.correlation}.

The window function used in this study is a semi-circle centered on each galaxy of radius $a$ in units of the its host halo's virial radius projected on the sky. This choice of window function is not circularly symmetric and it can be thought of the product of a top hat and a rectangular filter, $W(x,y)=W_1(x,y)W_2(x,y)$,
\begin{eqnarray}\label{eq:rksz1.appendix.aperturevariance.windowfunction}
    W_1(x,y) = 
    \begin{cases}
        \frac{1}{\pi a^2} & \text{if $x^2+y^2 \leq a^2$}\\
        0                 & \text{otherwise}
    \end{cases}\\
    W_2(x,y) =
    \begin{cases}
        1 & \text{if $|y|\leq a \wedge x \leq a \wedge x \geq 0$} \\
        0 & \text{otherwise}
    \end{cases}
\end{eqnarray}
for a right aperture.  For a left aperture, $W_2$ is displaced by $a$ to the left of the $x$-axis. The resulting half circle's window form, as a function of $\boldsymbol{\ell}\equiv(\ell_x,\ell_y)$ is $\widetilde{W} = \widetilde{W_1}\ast\widetilde{W_2}$, or
\begin{equation}\label{eq:rksz1.appendix.aperturevariance.analyticwindow}
    \widetilde{W}(\boldsymbol{\ell})=\frac{8}{a\ell}J_1\left(a\ell\right)
    \ast \frac{\sin\left(\frac{a\ell_x}{2}\right) \sin\left(a\ell_y\right)}{\ell_x \ell_y}\exp\left[\mp \mathrm{i}\frac{a\ell_x}{2}\right],
\end{equation}
here $J_1$ is the Bessel function of first kind, $\ast$ represents a convolution, the negative sign on the exponential corresponds to the right aperture and the positive sign to the left aperture. We assume a Gaussian beam function which depends on the CMB experiment's beam's full width at half maximum (FWHM), $b_{\ell}=\exp\left[-\frac{FWHM^2}{16\ln 2}\ell(\ell+1)\right]$. The power spectrum $C_{\ell}$ includes that of the CMB and any contributions of instrumental noise.

\section{Selecting galaxies from MaNGA for stacking.}\label{rksz1.appendix.manga}
We selected a set of galaxies for stacking by combining information from the MaNGA {\tt DRPALL} catalog with two value-added catalogs from SDSS DR15: the MaNGA Morphology Deep Learning DR15 and the {\tt GEMA-VAC}, see \S~\ref{rksz1.mangaplanck} for a brief description of them. The starting point are the 4,690 records with information in {\tt DRPALL}. Removing all records flagged with potential quality issues reduces the initial number to 4,196 (see \url{https://www.sdss.org/dr15/algorithms/bitmasks/} for a description of the bitmask used in the {\tt drp3qual} field).

We removed objects with more than one observation, that is, duplicates in the {\tt mangaid} field. There are 4,093 un-flagged objects with unique observations. Only objects in one of the three science target samples were considered, bringing the total number to 3,939. We computed their projected spin angle as described in \S~\ref{rksz1.mangaplanck.stack}, and removed those galaxies for which such a calculation yielded numerical errors, keeping 3,931 galaxies.

\begin{figure}\includegraphics[width=\linewidth]{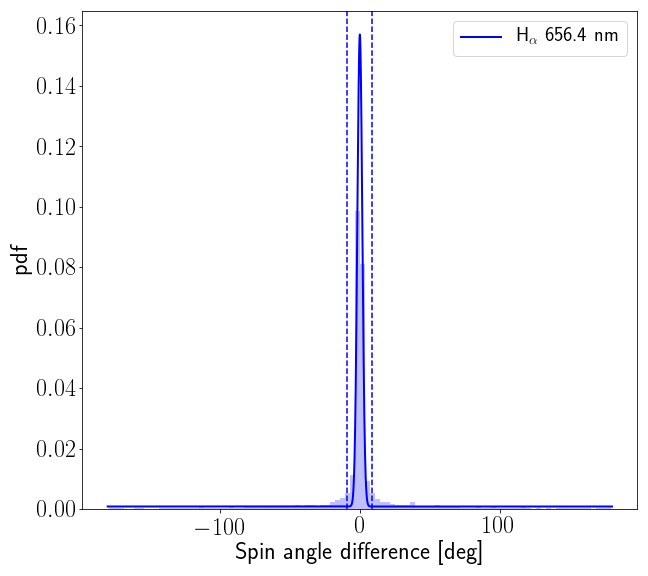}
    \caption{Histogram showing the difference between the spin angle computed using  the $H_{\alpha}$ line and the angle derived from O\,II (average of both O\,II lines). A mixture model with a Gaussian and a uniform component is a poor fit (indicated by the fat tails). To compensate for the badness of fit, we apply a 5$\sigma$ cut to identify outliers.}
    \label{fig:rksz1.appendix.manga.spin}
\end{figure}

A key assumption in the analysis of \textit{Planck} data on MaNGA galaxies is that the spin of the outer gaseous halo is aligned with that of the inner regions, which are the ones probed by MaNGA IFU spectrographs. We deemed this assumption more likely if the inner kinematics probed by different tracers are consistent with each other. To test for consistency, we also computed the spin angle using the H$_{\alpha}$ line (6,564\,\r{A}). We modeled the difference between this angle and the one from the O\,II line by a random variable whose pdf is a combination of a (zero mean) Gaussian and a uniform distributions, as is shown on Fig.~\ref{fig:rksz1.appendix.manga.spin}. The rationale for this choice is that, while most galaxies show a high correlation between spins estimated with different emission lines, some show little or no correlation. We calculated the best fit values for the Gaussian width and the uniform distribution height, and used the former to discard galaxies for which the difference between the spin angle estimated from O\,II and that from H$_{\alpha}$ exceeds five standard deviations of the Gaussian component. The standard deviation that maximizes the likelihood of our data is 1.8\,deg (although the fit is not good, see Fig.~\ref{fig:rksz1.appendix.manga.spin}), which indicates a very tight correlation between the spin angle measured using H$_{\alpha}$ and O\,II.  If that standard deviation is representative of the true uncertainty on the spin angle, our measurements will not be severely affected by errors in the spin angle estimation (see Appendix~\ref{rksz1.appendix.filtermisplacement}). A visual inspection of some of the outliers showed that they were either galaxies with a complex velocity field (i.e. no clear overall rotation pattern) or face-on systems that would contribute little to a rkSZ measurement. Removing the outliers shrank our stacking sample to 2,901 galaxies.

Adding morphological information allowed us to reject galaxies with a high probability of being interactive systems ({\tt P\_MERG}$>$0.95). This was motivated by the fact that the kinematics of the outer regions of interacting systems can be perturbed to the point of having little correlation with the inner kinematics probed by MaNGA. This further reduced the size of our stacking sample to a final number of 2,664 galaxies.

\end{document}